\documentclass[sigconf]{acmart}
\AtBeginDocument{%
  \providecommand\BibTeX{{%
    \normalfont B\kern-0.5em{\scshape i\kern-0.25em b}\kern-0.8em\TeX}}}
\copyrightyear{2022}
\acmYear{2022}
\setcopyright{acmcopyright}\acmConference[KDD '22]{Proceedings of the 28th ACM SIGKDD Conference on Knowledge Discovery and Data Mining}{August 14--18, 2022}{Washington, DC, USA}
\acmBooktitle{Proceedings of the 28th ACM SIGKDD Conference on Knowledge Discovery and Data Mining (KDD '22), August 14--18, 2022, Washington, DC, USA}
\acmPrice{15.00}
\acmDOI{10.1145/3534678.3539308}
\acmISBN{978-1-4503-9385-0/22/08}

\usepackage{amsmath,amsfonts}
\usepackage{subfigure} 
\usepackage{algorithm}
\usepackage{balance}
\usepackage{algorithmic}
\usepackage{bm}
\usepackage{graphicx}
\usepackage{textcomp}
\usepackage{xcolor}
\newcommand*{\innerproduct}[2]{%
    \if@display
        \left\langle #1,#2\right\rangle
    \else
        \langle #1,#2 \rangle
    \fi}
\begin{document}
\title{FedWalk: Communication Efficient Federated Unsupervised Node Embedding with Differential Privacy}

\author{Qiying Pan}
\affiliation{%
  \institution{Shanghai Jiao Tong University}
  \streetaddress{Dongchuan Road 800}
  \city{Shanghai}
  \state{}
  \country{China}
  \postcode{200240}
}
\email{sim10\_ arity@sjtu.edu.cn}

\author{Yifei Zhu}
\authornote{Corresponding author}

\affiliation{%
  \institution{Shanghai Jiao Tong University}
  \streetaddress{Dongchuan Road 800}
  \city{Shanghai}
  \state{}
  \country{China}
  \postcode{200240}
}
\email{yifei.zhu@sjtu.edu.cn}


\begin{abstract}
Node embedding aims to map nodes in the complex graph into low-dimensional representations.  The real-world large-scale graphs and difficulties of labeling motivate wide studies of unsupervised node embedding problems. Nevertheless, previous effort mostly operates in a centralized setting where a complete graph is given. 
With the growing awareness of data privacy, data holders who are only aware of one vertex and its neighbours demand greater privacy protection. 
In this paper, we introduce FedWalk, a random-walk-based unsupervised node embedding algorithm that operates in such a node-level visibility graph with raw graph information remaining locally.  FedWalk is designed to offer centralized competitive graph representation capability with data privacy protection and great communication efficiency. FedWalk instantiates the prevalent federated paradigm and contains three modules. We first design a hierarchical clustering tree (HCT) constructor to extract the structural feature of each node. A dynamic time warping algorithm seamlessly handles the structural heterogeneity across different nodes. Based on the constructed HCT, we then design a random walk generator, wherein a sequence encoder is designed to preserve privacy and a two-hop neighbor predictor is designed to save communication cost. The generated random walks are then used to update node embedding based on a  SkipGram model. Extensive experiments on two large graphs demonstrate that FedWalk achieves competitive representativeness as a centralized node embedding algorithm does with only up to  1.8\% Micro-F1 score and 4.4\% Marco-F1 score loss while reducing about 6.7 times of inter-device communication per walk.
\end{abstract}


\maketitle
\section{Introduction}

With the rise of graph convolutional networks and the abundant graph-structured data, graph representation learning has drawn great attention as it effectively extracts the structural information from complex graphs. 
Node embedding is a representation learning problem that captures the features of nodes and topological information in a graph with vectors.
The essence of node embedding is to represent each node such that the distance in the embedding space measures the dissimilarity among nodes in the original graph. 
If two nodes have similar structural features, the distance of their corresponding embedding vectors is small. Node embedding can be deployed in various downstream prediction tasks, such as link prediction \cite{wang2018shine}, anomaly detection \cite{yu2018netwalk}, and so on. 

Typical node embedding approaches follow a similar encoder-decoder framework. First, an encoder is defined to map a node to a vector. Then a node proximity function is defined to examine whether the current encoding result approximates the topological information of the original graph. Last but not least, the encoder is optimized based on the previous examination result. The key differences among various node embedding approaches are the definitions of encoders and node proximity functions. For unsupervised node embedding, node proximity functions typically rely on random walk methods due to their expressiveness and efficiency. Deepwalk \cite{perozzi2014deepwalk} pioneers the random-walk-based embedding by leveraging unbiased random walk to measure the similarity between each node. Node2vec \cite{grover2016node2vec} later replaces unbiased random walk with biased one to capture local microscopic and global macroscopic views of a graph. On the other hand, for supervised node embedding, node proximity functions are usually defined based on the labels in the downstream tasks. As neural networks have shown their great power in various fields, some frameworks also deploy neural networks to embed nodes in supervised and unsupervised settings \cite{kipf2016semi, hamilton2017inductive, velivckovic2017graph}. 


All the representation learning methods discussed above operate in a centralized setting, where a server owns the complete graph information. However, the growing awareness of privacy and the establishment of regulations and laws, e.g., GDPR \cite{EUdataregulations2018} and CCPA \cite{CAdata}, demand stricter data privacy protection. 
With the increasing computing power of mobile devices, a federated paradigm is proposed by Google and later being widely studied and deployed in real life\cite{mcmahan2017communication}. In the federated paradigm, devices with private data work collaboratively with a central server to conduct machine learning and data analytics tasks without uploading the raw data. The core of the federated paradigm is to train local models on individual devices first. Then the trained local models are aggregated at the server side to generate a global model after multiple iterations between clients and the server. In this whole process, only the extracted insights are uploaded to the server without revealing clients' raw data. 
The applications of federated learning on regular data structures, e.g., images \cite{liu2020fedvision}, text\cite{yang2018applied}, have demonstrated tremendous success in real life.


With the wide prevalence of graph-structured data, the federation of graph embedding, especially graph neural network, has also attracted significant attention. Existing federated graph embedding studies can be categorized into three types depending on how much graph information each local client owns: node-level\cite{mei2019sgnn,meng2021cross}, subgraph-level\cite{wu2021fedgnn} and graph-level\cite{zhou2020vertically,zheng2021asfgnn,peng2021differentially} federated graph embedding. 
Whereas graph-level and subgraph-level federated graph embedding still assume each client holds a correlated complete graph or a part of a global graph, node-level federated graph embedding emphasizes the privacy of individual nodes in one global graph. Each node in the graph is owned by one client and it is only aware of its neighboring situation.
The server only maintains the minimal node membership information. 
 Many real-world networks such as social networks \cite{zhou2008brief} and sensor networks \cite{yick2008wireless} naturally born with this setting. 
 It is straightforward to notice that node-level graph embedding is the most challenging task since it contains the least structure information on each local client.

Previous works on node embedding in federated node-level scenarios mainly concentrate on supervised settings with labels from downstream tasks.
SGNN\cite{mei2019sgnn} takes advantage of a degree list to compute node distance and manages to perform supervised node embedding under privacy constraints.  GraphFL\cite{wang2020graphfl}, motivated by meta learning, addresses semi-supervised node embedding in federated networks. Nevertheless, in reality, most data are unlabeled; manual labeling data is also expensive due to the large scale of real-world graphs. These all motivate the study of unsupervised node embedding in the federated environment, which has barely been studied yet. \textit{
Considering the huge demand in processing unlabeled graph data and the growing restrictions on node-level data access, we aim at filling this gap in this paper. }



The challenges for providing unsupervised node embedding in the node-level federated setting come from three perspectives. First, \textit{learning}. Inaccessibility of graph information impedes deployment of conventional node proximity functions in federated graphs. With each client possessing just one-node information, achieving centralized competitive representation performance calls for new designs.
Second, \textit{privacy}. How to protect raw graph connection information from other clients and the central server, preferably theoretically guaranteed, is non-trivial. Naive deployment of the differential privacy measures may greatly affect the expressiveness of the learned results. Last but not least, \textit{communication efficiency.} The node size of a practical network can scale to tens of thousands or even larger. Straightforward collaborations among all clients incur prohibitive communication costs in practice and hinder its deployment. How to embed nodes communicationally efficiently should be seriously addressed. 


In this paper, we propose a federated random-walk-based node embedding approach, named FedWalk, that successfully operates in decentralized node-level visibility graphs. 
To do so, FedWalk first adopts a novel hierarchical clustering tree (HCT) constructor which hierarchically clusters vertices based on their linkage similarity in a federated graph. 
Second, it has a random walk sequence encoder to output a sequence with $\epsilon-$differential privacy. Last but not least, FedWalk includes a two-hop neighbor predictor to predict the possible two-hop nodes without actually visiting the direct ones. 
To the best of our knowledge, FedWalk is the first work for node-level federated unsupervised graph embedding, where each node is isolated and maintained by a data holder with only one-hop connection information available.
In summary, our contributions are summarized as follows:
\begin{itemize}
    \item We propose a novel federated unsupervised node embedding framework that robustly captures graph structures, guarantees differential privacy and operates with great communication efficiency.
    \item We propose a differentially private HCT constructor that extracts structural information from decentralized data holders. A dynamic time warping algorithm seamlessly handles the structural heterogeneity across different nodes.
    \item We propose a novel random walk generator with a differential private sequence encoder and a neighbor predictor for communication efficiency. 
    \item We prove that the dissimilarity value loss of node pairs in our framework is bounded. We also theoretically quantify the reduced communication cost benefited from our neighbor predictor.
    \item Extensive experiments on two datasets of different scales demonstrate that FedWalk can achieve centralized competitive results by losing up to 1.8\% Micro-F1 score and 4.4\% Marco-F1 score while decreasing about 6.7 times of inter-device communication per random walk. 
\end{itemize}


\section{Related Work}
\label{sec:rw}
In this section, we review the related works of this paper from three perspectives: unsupervised node embedding in the centralized setting; the emerging federated learning and analytics paradigm; the growing works on federated graph representation.
\subsection{Unsupervised Node Embedding}
Node embedding uses low-dimensional vectors to represent nodes in a graph. In practice, the scale of a graph is usually too big to have all nodes labeled. The inevitable absence of supervision information triggers the formation and development of unsupervised node embedding algorithms.  We can classify these algorithms into two sorts: (1) shallow networks with one hidden layer; (2) deep networks with multiple graph-based layers.  

Random-walk-based shallow embedding supports effective solutions for many graph-related tasks\cite{goyal2018graph}. DeepWalk\cite{perozzi2014deepwalk} pioneers this branch by 
treating the path of random walk similar to the word sequences processing in the NLP field and takes advantage of the Skipgram model to accomplish embedding. Since DeepWalk, we have witnessed significant efforts in random-walk-based shallow embedding, including Node2Vec\cite{grover2016node2vec}, HARP\cite{chen2018harp}, etc.

Deep embedding leverages multi-layer convolutional networks to learn graph representations. 
Application of convolutional networks in node embedding is started with GCN\cite{kipf2016semi} propagating neighboring embedding through layers. Since then, extensive efforts \cite{hamilton2017inductive} \cite{velivckovic2017graph} 
have been made to improve representation performance in practical large graphs.
Albeit the above well-developed node embedding methods, all of them require upload of raw data to the central server, which gradually becomes hard due to laws and regulations.

\subsection{Federated Learning and Analytics}
Federated learning and analytics have attracted great interest from both academia and industry due to their capability in conducting learning and analytic tasks without centralizing the raw data\cite{mcmahan2017communication}. Specifically, federated learning has been widely studied for collaborative neural network training. 
Tremendous efforts have been devoted to improving federated learning in terms of convergence speed, training accuracy, communication efficiency, fairness, etc. Several good surveys are presented here for interested readers\cite{li2020federated}.   
In practice, federated learning has also been applied in natural language processing\cite{yang2018applied}, computer vision\cite{liu2020fedvision},  and many other areas. 

Federated analytics, a recently proposed sibling problem, extends the applications of federated paradigm from neural network training to diverse data analytic tasks or unsupervised learning scenarios\cite{ramage_2020}.
TriHH\cite{zhu2020federated} discovers the heavy hitters in a population of data without uploading the data to the server. Federated location heatmap from Google applies distributed differential privacy to generate location heatmaps\cite{bagdasaryan2021sparse}. FedFPM \cite{zibo} presents a unified framework for frequent pattern analysis.
While most federated designs follow an iterative collaboration pattern, the exact extracted insights and aggregation methods vary greatly across different problems. This work pushes the boundary of the federated applications by studying a novel federated unsupervised node embedding problem. FedWalk also encompasses its design to provide communicating efficient, differentially private, centralized competitive node embedding.  

\subsection{Federated Learning on Graphs}
Unlike classical federated learning problems, the explicit linkage between nodes acts as a barrier to adapting algorithms in centralized graphs to federated ones. Due to privacy concerns, graphs cannot be uploaded to the server and can only be stored locally on each data holders. Despite this, some studies still succeed in designing algorithms to compute node embedding in federated graphs, especially under the subgraph-level visibility\cite{wu2021fedgnn} and graph-level visibility\cite{zhou2020vertically}\cite{zheng2021asfgnn}\cite{peng2021differentially}, where each client owns a part of the graph or one correlated graph.  In the most strict node-level federated graph works, devices only store the feature and edge information of one node. Leakage of node feature and edge information is forbidden. Under some circumstances, even the sharing of node labels is limited. This setting is applicable in social networks, Internet of Things (IoT) networks, and so on. Some studies have addressed the problem of inaccessibility of node data. SGNN\cite{mei2019sgnn} uses one-hot encoding to process local information and introduces a new node distance calculation method based on an ordered degree list to process structural information. A paradigm called CNFGNN\cite{meng2021cross} alternatively trains models on local devices and the server in a graph representing spatio-temporal data structure. 

Although there are some novel methods to train federated GNN, their models can only be applied in supervised settings with nodes labeled. How to provide node embedding in unsupervised federated scenarios remains to be solved. Furthermore, the extensive communication cost in federated node-level graph embedding should be carefully addressed to be practical in real-world systems.

\section{Preliminaries}
\label{sec:pre}
In this section, we first present a basic introduction to the classical node embedding process in the central situation. 
Then considering the special emphasis of privacy in a federated system, we introduce the preliminaries on a widely-used privacy protection measure, differential privacy. 

\subsection{Classical Node Embedding}
Node embedding is one of the most fundamental problems in graph theory.
For a graph $G=(V,E)$, node embedding represents the structural information of vertices. Vertices that are linked with shorter distances or share similar topological structures tend to be embedded with shorter distances. 

The node embedding algorithm is generally composed of three parts: (1) an encoder; (2) a dissimilarity function; (3) an optimizer. The encoder (ENC) is a function that maps a vertex to a fixed-dimension vector defined as Eq.\ref{eq:enc}.  
\begin{equation}
    ENC:V\rightarrow \mathbb{R}^d
    \label{eq:enc}
\end{equation}
One of the most basic encoders is shallow encoding which serves as a look-up table for all the vertices. A matrix stores all the embedding vectors and an indicator vector assigns different embedding vectors to different vertices. 

The dissimilarity function (DISSIM) in the form of Eq. \ref{eq:sim} defines to what extent two vertices are dissimilar to each other. 
\begin{equation}
\label{eq:sim}
    DISSIM:V\times V\rightarrow \mathbb{R}
\end{equation}
A typical dissimilar function is the probability of two vertices that don't co-occur in a random walk.  

The optimizer is a machine learning algorithm deployed to optimize the parameters in the encoder aiming to realize that the distance between two embedding vectors approximates the corresponding similarity value. The goal can be formulated as 
\begin{equation}
    \label{eq:goal}
    \min_{ENC} \sum _{v_1,v_2\in V}\text{dist}(ENC(v_1),ENC(v_2))- DISSIM(v_1,v_2).
\end{equation}

In unsupervised settings, random-walk-based graph embedding is widely used to learn vector representations for nodes. 
A random walk generator randomly samples the neighbors of the last vertex to form a sequence of vertices of a certain length. Using the embeded node proximity information in these sequences, models like the Skipgram model map the results of the random walks into real values representing certain node proximity. The probability of co-occurrence of two vertices in a random walk sequence represents the similarity between two vertices. 

\subsection{Differential Privacy}
Differential privacy (DP) \cite{dwork2006differential} is a prevalent measure defining whether an information-sharing strategy preserves privacy or not. It is motivated by the idea that if a single element in the dataset is replaced, the result of published information differs very small from the ideal one. The formal definition is shown in Def.\ref{def:dp}. 
\begin{definition}
\label{def:dp}
Let $\epsilon\in \mathbb{R}^+$ and $\mathcal{A}$ be a randomized algorithm. The algorithm $\mathcal{A}$ is defined as $\epsilon-$ differential privacy if for all datasets $D_1 $ and $D_2$ whose elements are the same except for one element and all sets $S\subseteq \text{Range}(\mathcal{A})$, 
\begin{equation}
    \text{Pr}[\mathcal{A}(D_1)\in S]\leq \exp{(\epsilon)}\text{Pr}[\mathcal{A}(D_2)\in S]
\end{equation}
\end{definition}
Many DP mechanisms have been  proposed, such as Laplace mechanism \cite{dwork2006calibrating}, exponential mechanism\cite{mcsherry2007mechanism} and randomized response\cite{warner1965randomized}. 

\section{Design}
In this section, we first briefly introduce the overall design of our framework, FedWalk. We then present two core components of FedWalk, hierarchical clustering tree constructor and random walk generator in detail. 
\label{sec:design}
\subsection{Design Overview}
\begin{figure}[H]
    \centering
    \includegraphics[width=0.48\textwidth]{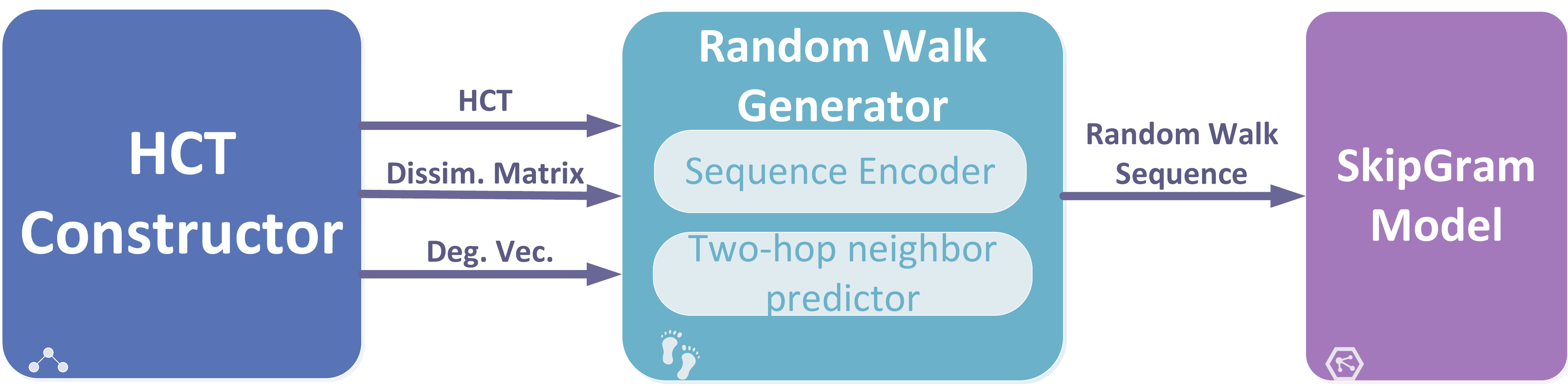}
    \caption{FedWalk framework}
    \label{fig:fedwalk}
\end{figure}
FedWalk is a random-walk-based graph embedding algorithm for unsupervised federated node embedding when each client \footnote{Clients and data holders are used interchangeable in this paper.} only has its neighboring information. The server is not allowed to access the raw local graph information and is only aware of which node is in the graph.
FedWalk can not only embeds structure information without labeled data but also provides differential privacy to the clients' local connection information. 
FedWalk can be divided into three parts as shown in Fig. \ref{fig:fedwalk}.
To tackle the problem of inaccessibility of global graph, we introduce a differentially private \textit{hierarchical clustering tree (HCT) constructor} based on a newly-proposed distance computing metric that clusters vertices according to their structural information. The tree is constructed gradually among clients under the coordination of a central server. 
A revised \textit{random walk generator} is then developed to construct random walk sequences in federated settings based on the constructed hierarchical clustering tree. 
In the random walk generator, FedWalk contains a sequence encoder to encrypt the edge information to provide privacy and a two-hop neighbor predictor to save the communication cost among devices. 
The typical \textit{Skip-Gram model} is then used to improve node embedding based on the random walk sequence generated in the second part. 



\subsection{Hierarchical Clustering Tree Constructor}
The HCT constructor aims to build a hierarchical clustering tree in federated settings which will later be used for encoding and transmission of random walk sequences. It is designed to preserve structural information with privacy guaranteed. 

\begin{figure}[h]
    \centering
    \subfigure[Original Graph]
    {
        \includegraphics[width=0.22\textwidth]{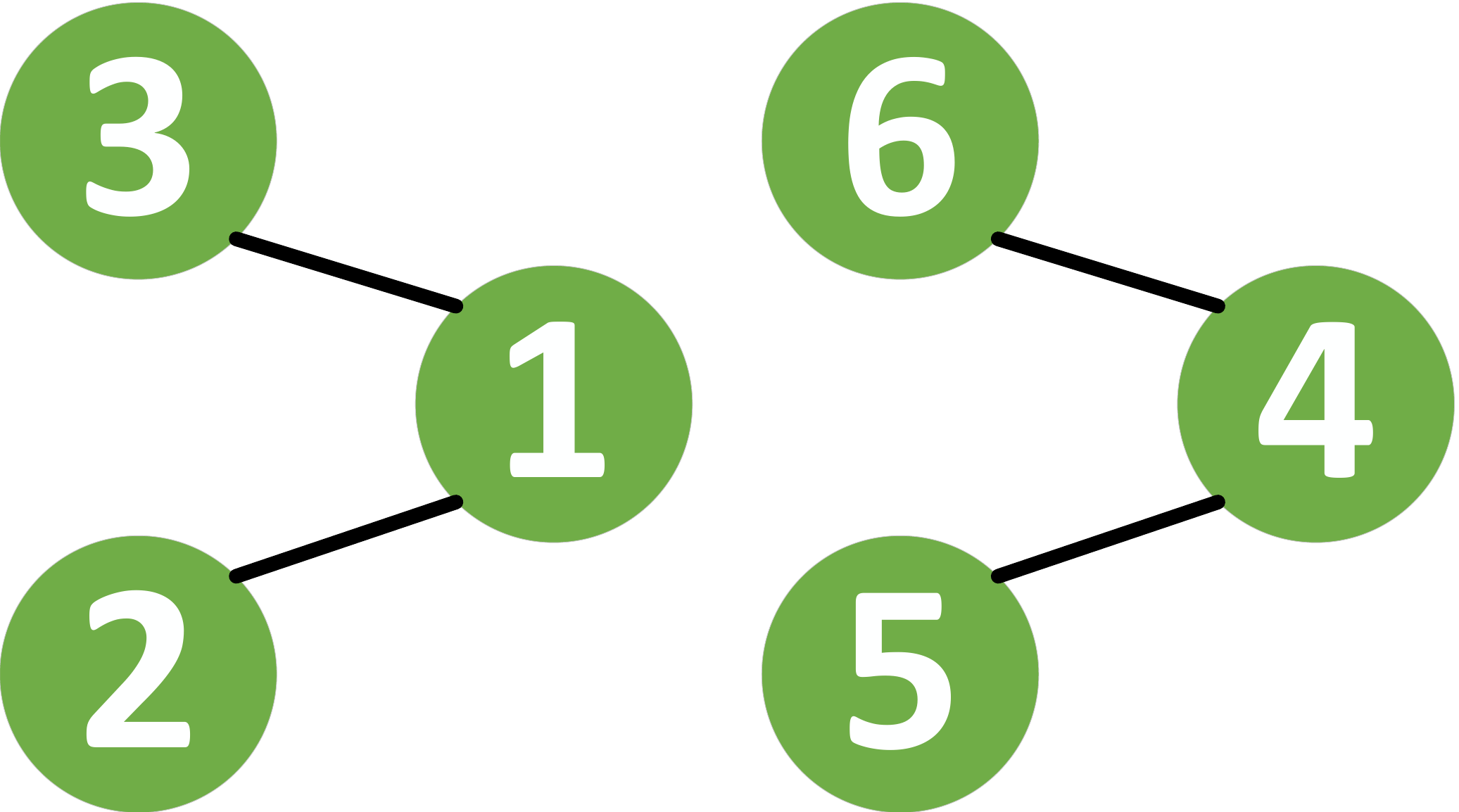}
      
    }
    \subfigure[Corresponding HCT]
    {
        \includegraphics[width=0.22\textwidth]{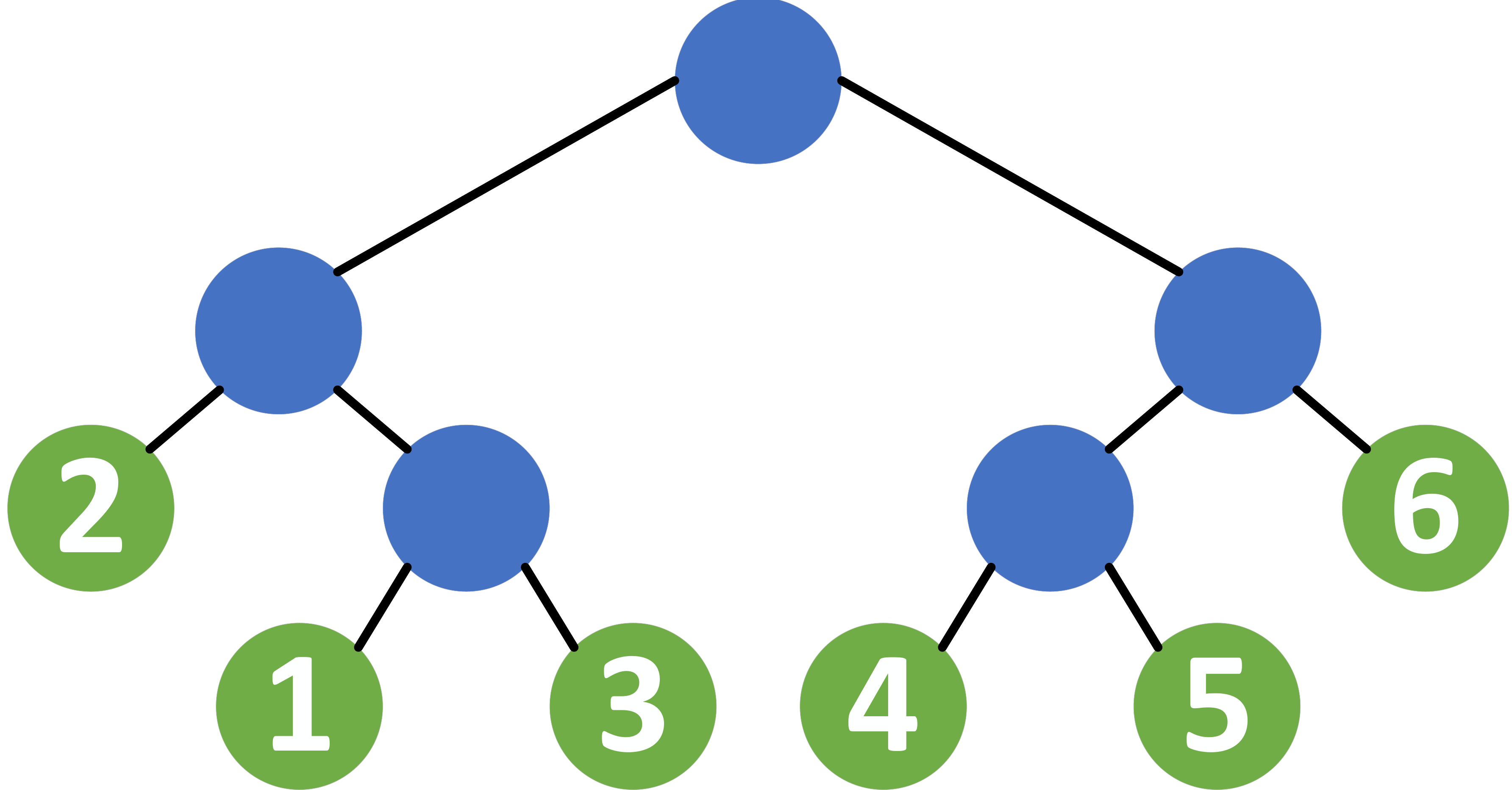}
  
    }
    \caption{An example of HCT}
    \label{fig:hct}
\end{figure}

A hierarchical clustering tree (HCT) clusters vertices in the graph using edge information. It is widely used in NLP tasks\cite{ushioda1996hierarchical}, biological information processing\cite{yin2014measure} and so on. A hierarchical clustering structure separates vertices into two groups and then these groups further split themselves into two small groups. This splitting process continues until every group only contains one vertex.  The leaf nodes in the tree correspond to the vertices in the graph. A none-leaf node, sometimes called an  internal node, clusters a group of vertices by putting them to its left subtree or right subtree. As Figure \ref{fig:hct} illustrates, the root node clusters the vertices into two groups $\{ 1,2,3\}$ and $\{ 4,5,6\}$. The internal nodes further split the groups into different clusters. It is straightforward to observe that every HCT has $|V|-1$ internal nodes for a certain graph $G=(V,E)$.


In a centralized graph, a typical framework to construct a hierarchical clustering tree is first to compute the structural dissimilarity matrix (e.g., topological overlapping matrix \cite{ravasz2002hierarchical}) between all vertices pairs of the graph. 
Then it utilizes a  hierarchical clustering algorithm to build a tree based on the dissimilarity matrix. However, since the federated settings forbid direct update of edge information to the server, we need to design a new approach to compute the dissimilarity matrix.

In our design, the constructor first computes an ordered degree matrix for each vertex to represent its structural feature. Dynamic Time Warping (DTW) is introduced to compute the dissimilarity between each vertex pair due to the heterogeneous sizes of the matrices  emerged in our setting. Then these dissimilarity values help to form a hierarchical clustering tree. 
The process is illustrated in Fig.\ref{fig:hctcon}. The complete algorithm for HCT construction is displayed in Alg. \ref{hct-constr} in Appendix B due to the space limit.
We then present the design details as follows. 
\begin{figure}[h]
    \centering
    \includegraphics[width=0.48\textwidth]{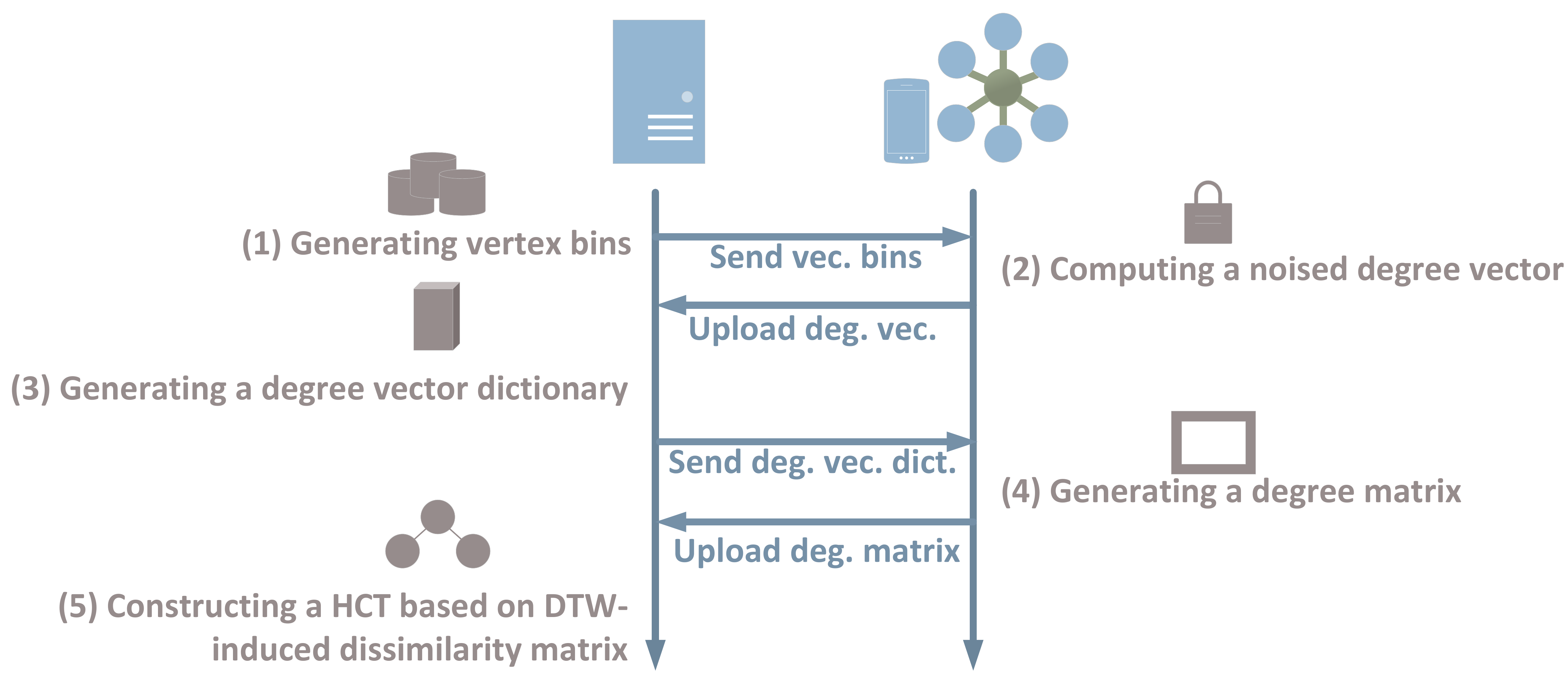}
    \caption{HCT constructor framework}
    \label{fig:hctcon}
\end{figure}


\paragraph{Ordered Degree matrix} We first define an ordered degree matrix for each vertex to compute the dissimilarity between two different nodes. 
The server randomly numbers the vertices in its system and virtually groups all the vertices into $k$ bins. Each bin contains at least one vertex. Larger $k$ preserves finer structural information while smaller $k$ preserves more privacy. In our scenario, $k$ is upper-bounded by $\ln{|V|}$ due to privacy concerns. The server then sends the group plan to each device. Since every device stores a subgraph representing the link relationship between itself and neighboring devices, each device can count the number $c_i,i\in [1,2,\cdots , k]$ of its neighbors in the $i$-th bin. $c_i$ values are then noised with Laplacian noise $Lap(0,\frac{1}{\epsilon})$ to guarantee $\epsilon-$ differential privacy, forming a vector $\bm{c_v}=(c'_1,c'_2,\cdots ,c'_k)\in \mathbb{R}^k$ for a vertex $v\in V$. It is worth mentioning that elements in $\bm{c_v}$ can be negative real numbers due to the Laplacian noise. Moreover, the sum of vector elements $\sum_{i=1}^kc'_i$ is the
noised degree value of vertex $v$. 
3
These vectors $\bm{c_{v_1}},\bm{c_{v_2}},\cdots, \bm{c_{v_n}}$ are sent to the server and the server advertises these vectors to all the devices. 
Each device corresponds to a vertex $v$ with degree \textit{deg(v)} in the graph. Denote the neighbors of a node $v$ as $u_1, u_2, \cdots, u_{deg(v)}$.  Each device  organize its neighboring vertices into an ordered list $ N(v)$, where $N(v)=\left[ u_{(1)},u_{(2)},\cdots ,  u_{(deg(v))}\right]$, in an ascending order based on the estimated degrees of its neighboring vertices. 
In this way, each device can form an ordered degree matrix $\bm{M_v}$ (Eq.\ref{degree_matrix}) of size $|N(v)|\times k$ where each row is the degree vector of the corresponding neighboring vertex. 
\begin{equation}
\label{degree_matrix}
    \bm{M_v}=\left[ 
    \begin{aligned}
    &\quad\bm{c_{u_1}}\\
    &\quad\bm{c_{u_2}}\\
    &\quad\quad\vdots \\
    &\bm{c_{u_{|N(v)|}}}\\
    \end{aligned}\right]
\end{equation}

This ordered degree matrix represents 2-hop structural information for each vertex. Metrics from different devices have different numbers of rows because they have different numbers of neighbors. Each device uploads this ordered matrix to the server to calculate the dissimilarity matrix.

\paragraph{Dissimilarity matrix} 
Given the structure-embedded ordered degree matrix from the previous step, the server constructs a dissimilarity matrix for the whole graph. 
A dissimilarity matrix in a graph of size $|V|\times |V|$ quantifies dissimilarity for each node pair. Since the ordered degree matrix is expressive of the structural information of a vertex, we use the distance between two ordered degree matrices to determine the dissimilarity matrix.  However, the metrics of each device can have unequal numbers of rows because they have different numbers of neighbors. Classical l1-norm of the subtraction of two metrics no longer applies. Therefore, we adopt Dynamic Time Warping (DTW)\cite{berndt1994using} which computes the distance between two sequences of unequal length by deploying the dynamic programming algorithm. The algorithm can be further extended to compute distances of arrays of different dimensions. 

Let us suppose that two matrices $\bm{M_u}$ and $\bm{M_v}$ have $x$ and $y$ rows where $x\ge y$ correspondingly. As mentioned before, the first matrix represents neighboring vertices $N(u)$ while the second matrix represents neighboring vertices $N(v)$. Each row vector can be seen as a feature vector of a vertex. The server uses l1-norm to compute the dissimilarity between two row vectors, which aims to pair each vertex in $N(u)=[u_1,u_2,\cdots ,u_x]$ with exactly one vertex in $N(v)=[v_1,v_2,\cdots ,v_y]$ with smallest dissimilarity. This can be achieved using dynamic programming formulated in Eq. \ref{dp} 
\begin{equation}
    \label{dp}
    \text{cost}(i,j)=
    \begin{aligned}
    \\
    &\min \{\text{cost}(i-1 ,j),
\text{cost}(i,j-1), \text{cost}(i-1,j-1)\}\\
&+||\bm{c_{u_i}}-\bm{c_{v_j}}||_1, \forall i\in [1,x], j \in [1,y]
\end{aligned}
\end{equation}
where cost is a function measuring how well the former $i$ vertices in $N(u)$ are paired with the former $j$ vertices in $N(v)$. 
\begin{equation}
    \text{dissim}(u,v)=\text{cost}(x,y)
    \label{dissim}
\end{equation} is the dissimilarity value for two vertices $u$ and $v$. The dissimilarity value is noised due to privacy restrictions but it is proved to have upper-bounded information loss according to the following Theorem \ref{th:ub}. Proofs of our theorems are presented in the appendix.

\begin{theorem}
\label{th:ub}
The expectation $\mathbb{E}[\text{dissim}(u,v)-\text{dissim}'(u,v)]$ where $\text{dissim}'(u,v)$ is the original dissimilarity value of vertices $u$ and $v$ is bounded by 
\begin{equation}
    \mathbb{E}[\text{dissim}(u,v)-\text{dissim}'(u,v)]\leq \frac{3k(\max_{v\in V}|N(v)|)^2}{2\epsilon}.
    \label{eq:ub}
\end{equation}
\end{theorem}
In practice, the left-hand side in Eq.\ref{eq:ub} is dependent upon the degrees of two vertices.

The server computes the dissimilarity for every vertex pair in this way and gets a dissimilarity matrix. Then it generates an HCT using a hierarchical algorithm based on this dissimilarity matrix. Considering the large scale of practical networks, we use hybrid hierarchical clustering \cite{sun2009efficient} to construct the HCT. Other compatible clustering algorithms can also be used in our framework. 

\subsection{Random Walk Generator}
Based on the constructed HCT, we design a random walk generator to generate random walk of limited length $l$ starting with a specific vertex. The server sends each vertex a tuple $(l,\epsilon, p)$ to define and start a random walk. Value $\epsilon$ is the noise parameter to encode vertex with quantifiable privacy level. Value $p$ is the probability of whether to use a 2-hop neighbor predictor, acting like a knob on controlling communication cost. As long as receiving the tuple, the corresponding device randomly selects a vertex from its neighbors. It utilizes the random walk sequence encoder to encode its vertex index and appends the encoder result to the tail of the walk sequence. Then it decides whether to trigger the 2-hop neighbor predictor based on the probability $p$. The length $l$ is assumed to be greater than 2 to capture enough graph structures.
If the predictor is triggered, it encodes the selected 2-hop neighbor vertex index and appends the encoded index to the output sequence as well. 
It sends the tuple $(l-2, \epsilon ,p)$ and the current walk sequence to the predicted 2-hop neighbor vertex to continue the walk. If the predictor is not triggered, it directly sends the tuple $(l-1, \epsilon ,p)$ and the current walk sequence ended to the selected neighbor vertex. The random walk stops when the first number received in the tuple is 1. The corresponding device, also representing the ending vertex of the whole random walk, sends the encoded random walk sequence to the server. The complete algorithms are presented in Alg.\ref{random-walk-gen}. The detailed procedures for random walk sequence encoder and two-hop neighbor predictor are presented as follows.

\paragraph{Random walk sequence encoder}

The random walk sequence encoder takes advantage of exponential mechanism to noise the elements in the sequence to protect the privacy of edge information. 

Exponential mechanism deals with contradiction between information loss and privacy protection through a score function $u:D\times \text{Range}(\mathcal{A})\rightarrow \mathbb{R}$, which measures how informative the output of $\mathcal{A}$ is given $d\in D$.  The mechanism assigns a sampling probability defined in Eq. \ref{em} to the element in the range of $\mathcal{A}$ given different input $d\in D$ and samples an output based on the probability.  
\begin{equation}
    \label{em}
    \text{Pr}[r|d] = \frac{\exp (\epsilon u(d,r))\mu (r)}{\int \exp (\epsilon u(d,r))\mu (r)dr}
\end{equation}
where $\mu (r)$ is the distribution of $r$ in Range of $\mathcal{A}$. 
\begin{algorithm}[h]
\caption{Random Walk Generator}
\label{random-walk-gen}
\begin{algorithmic}[1]
\REQUIRE a HCT $T$; dissimilarity matrix; the length of the walk $l$; vertices $V$; noise parameter $\epsilon$; triggering probability $p$; number of walks per vertex $\gamma$;\\
\textbf{Server operation}\\
\STATE Send $T$ and dissimilarity matrix to all devices; \\
\FOR{$i=1\rightarrow \gamma $}
\FOR{each device}
\STATE Server sends a tuple $(l,\epsilon ,p)$ to the local device;\\
\STATE Execute \textbf{Device operation};\\
\STATE Server receives the random walk sequence;
\STATE Server inputs the random walk sequence to the SkipGram Model;\\
\ENDFOR
\ENDFOR

\quad \\
\textbf{Device operation}\\
\IF{the random walk sequence is empty}
\STATE{Initiate a new random walk sequence with an empty array}
\ENDIF
\STATE Randomly select a vertex $u$ from its neighbors;\\
\STATE Encode the vertex index according to Eq.\ref{sp};\\
\STATE Append the encoded index to the end of the random walk sequence;\\
\IF{$l=1$}
\STATE Send the random walk sequence to the server; \\
\ELSIF{$l>2$}
\STATE Randomize a real number from $[0,1)$ ;\\
\IF{the randomized number$<p$}
\STATE Choose the row $\bm{c_u}$ which corresponds to the degree vector of $u$ from the degree vector dictionary;\\
\STATE Initiate a vertex pool with an empty set;\\
\FOR{bin $i$ in bins}
\STATE Add $\bm{c_u}[i]$ vertices which are nearest to $u$ in tree $T$ to the vertex pool;\\
\ENDFOR
\STATE Randomly pick a vertex $u'$ from the pool;\\
\STATE Encode the vertex index $u$ according to Eq.\ref{sp};\\
\STATE Append the encoded index of $u$ to the end of the random walk sequence;\\
\STATE Device sends a tuple $(l-2,\epsilon,p)$ and the current random walk sequence to device $u'$;\\
\STATE Device $u'$ execute \textbf{Device operation};\\
\RETURN
\ENDIF
\ENDIF
\STATE Device sends a tuple $(l-1,\epsilon,p)$ and the current random walk sequence to device $u$;\\
\STATE Device $u$ execute \textbf{Device operation};\\
\end{algorithmic}
\end{algorithm}

The advantage of exponential mechanism is the ensuring of the utility of encoding result due to its dependence upon the score function. Another advantage is that it provides a way to encode discrete datasets. In our scenario, the random walk sequence should still imply the structural information of the graph despite alternation of the original sequence. In addition, the dataset $V$ to encrypt is a discrete vertex set. Hence, we choose exponential mechanism to preserve $\epsilon-$ differential privacy. 

We define the score function to be used in Eq. \ref{em}  as following: 
\begin{equation}
    \label{sf}
    u(v_1,v_2)= -\text{dissim}(v_1,v_2)|\text{leaves}(T[v_1\lor v_2])|
\end{equation}
where dissim function is defined in Eq.\ref{dissim}, $T$ is the subtree rooted by the least common ancestor of $v_1$ and $v_2$ nodes in HCT constructed before and leaves function returns all the leaves of a subtree. For two vertices with similar structural information, they tend to have a smaller dissimilarity value and are put more closely in the HCT. If two leaves are close to each other in a tree, $T[v_1\lor v_2]$ have very few leaves. Hence, Eq.\ref{sf} measures how similar two vertices are in terms of structure, which implies how informative the random walk sequence is if the original vertex is substituted with the other. 

It is obvious that any vertex $v\in V$ follows a uniform distribution 
\begin{equation}
\label{pr}
    Pr[v]=\frac{1}{|V|}.
\end{equation}
Hence, after we plug Eq.\ref{sf} and Eq.\ref{pr} into Eq.\ref{em}, the sampling probability becomes
\begin{equation}
    \label{sp}
    \text{Pr}[v_2|v_1] = \frac{\exp (-\epsilon \text{dissim}(v_1,v_2)|\text{leaves}(T[v_1\lor v_2]))}{\sum_{v_2\in V} \exp (-\epsilon \text{dissim}(v_1,v_2)|\text{leaves}(T[v_1\lor v_2]))}
\end{equation}
Using Eq.\ref{sp}, we encode the random walk sequence by sampling one vertex given the original index. 
\paragraph{Two-hop neighbor predictor}
To decrease inter-device communication cost, we design a two-hop neighbor predictor to directly build a bridge from the $i$th element in the sequence to the $(i+2)$th element, saving the communication with the $(i+1)$th element. Our predictor takes advantage of HCT and the ordered degree matrix $\bm{M_v}$ obtained in the HCT constructor. The predictor is triggered given probability $p$ customized by the server.

Each device is aware of the vectors $\bm{c_{u_1}}, \bm{c_{u_2}},\cdots ,\bm{c_{|N(v)|}}$ with respect to its 1-hop neighbors. The device has randomly selected one neighbor denoted as $u_i$ before. In HCT constructor, the degree vector $\bm{c_{u_i}}=(c_1',c_2',\cdots ,c_k')$ of $u_i$ provides the number of its 2-hop neighbor connected to $u_i$ in $k$ different bins. Since HCT hierarchically clusters vertices according to their linkage information, for bin $j$, the predictor selects $c_j'$ vertices that are closest to the $u_i$ in HCT. These selected vertices form a pool. Then a vertex is uniformly drawn from this pool as the predicted 2-hop neighbor of $v$. In this way, the device can directly communicate with the device with the predicted 2-hop neighbor index. This reduces the times of communication by saving the contact involving the 1-hop neighbor $u_i$. We formally analyze the saved communication cost caused by our predictor and present the theorem as follows. 
\begin{theorem}
For a $l-$long random walk sequence, it is expected that 2-hop neighbor predictor will reduce $|V|\gamma \mathcal{A}$ device to device communication given the triggering probability $p$ and the number of walks per vertex $\gamma $ where 
\begin{equation}
    \mathcal{A}=l-1-\left[ \frac{l-2}{1+p}+(2-2p-\frac{1}{1+p})\frac{1-(-p)^{l-2}}{1+p}\right]
\end{equation}
\label{probcomm}
\end{theorem}
As long as a random walk sequence is generated, the server computes the probability of co-occurrence for any vertex pair in this sequence, which provides a measurement of the similarity of two vertices in a graph. A Skipgram model is used to update node embedding based on the computed probability such that the distance between node embedding vectors approximates the corresponding vertex similarity. 

\section{Experiments}
In this section, we present the empirical results and analysis of our proposed algorithm. We first examine how FedWalk performs compared with the existing baseline in traditional node embedding tasks. Then we discuss the influence of key hyper-parameters upon the model performance.  
\label{sec:exp}
\subsection{Experiment Setup}
The performance of FedWalk is measured through multi-label classification tasks. First, we deploy FedWalk to embed the vertices. Then, we randomly split the two datasets into a training dataset and a testing dataset based on a proportion $T_R$. $T_R$ is the proportion of training data in full data. We set the vertex embedding vectors of the training data as features $\bm{x}$ and the labels of the training data as targets $\bm{y}$ and feed the $(\bm{x}, \bm{y})$ pairs into a one-vs-rest logistic regression. $\bm{y}$ is the indicator vector of the multiple labels for one vertex. Last we test the regression model using testing data. 

We evaluate the performance of FedWalk based on two social network datasets:  Blogcatalog\cite{tang2009relational} and Flickr\cite{tang2009relational}. Specifically, 
\begin{itemize}
    \item Blogcatalog dataset is crawled from a blog posting website Blogcatalog\footnote{ http://www.blogcatalog.com} representing the online social relationship between bloggers. The vertex labels are the interest of the bloggers. 
    \item Flickr dataset is crawled from an image sharing website Flickr\footnote{http://www.Flickr.com }. The vertices are the users and the edges represent that users add others to their contact lists. The labels are the interest groups of the users.
\end{itemize}
Blogcatalog dataset has 10312 vertices, 333983 edges, and 39 labels; Flickr dataset has  80513 vertices, 5899882 edges, and 195 labels. 
Two graphs are of different sizes so that the scalability of our algorithms can be examined.

Since we are the first to study the node-level federated unsupervised embedding problems, and the existing federated works cannot be applied in our scenario, we compare our FedWalk with DeepWalk\cite{perozzi2014deepwalk}, a classical node embedding method in centralized settings. It regards the random walk sequences as the word sequences and deploys Skipgram model to embed the nodes with vectors. In our experiments, both methods embed vertices with walk length $l=40$, window size $w=10$, embedding dimension $d=128$ and $\gamma =80$ walks per vertex. For FedWalk, we set the number of bins $k=\lfloor \ln |V| \rfloor$, the noise parameter $\epsilon=2$ and the probability of triggering 2-hop neighbor predictor $p=0.2$. The experiments are done with an Intel Core i7-10700 CPU clocked at 2.90GHz, an 8GB-memory NVIDIA RTX 3070, and 16GB memory. 

To quantify the performance of multi-label classification, we use macro-F1 and micro-F1 to measure its accuracy. 
\begin{itemize}
\item Macro-F1 is the average F1 score of all labels. 
\begin{equation}
    Macro-F1 = \frac{\sum _{l\in L}F1(l)}{|L|}
\end{equation}
\item Micro-F1 is calculated by counting the overall true positives, false negatives and false negatives for all labels. 
\begin{equation}
    Micro-F1 = \frac{2PR}{P+R}
\end{equation}
where $P=\frac{\sum _{l\in L}TP(l)}{\sum _{l\in L}TP(l)+FP(l)}$ and $R=\frac{\sum _{l\in L}TP(l)}{\sum _{l\in L}TP(l)+FN(l)}$. 
\end{itemize}
\subsection{Results}

\begin{figure}[h]
    \centering
    \subfigure[Micro-F1 score]
    {
        \includegraphics[width=0.22\textwidth]{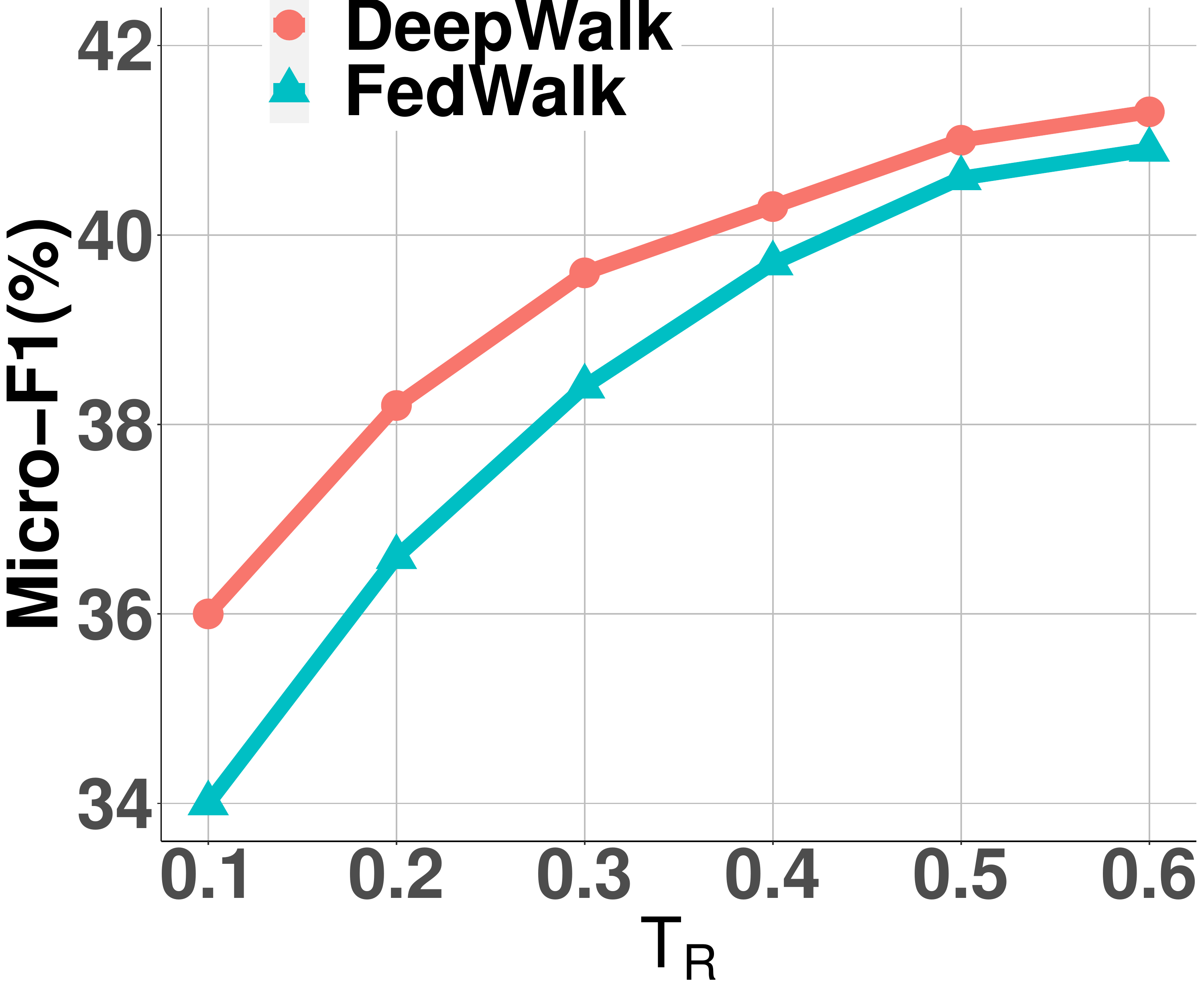}
      
    }
    \subfigure[Macro-F1 score]
    {
        \includegraphics[width=0.22\textwidth]{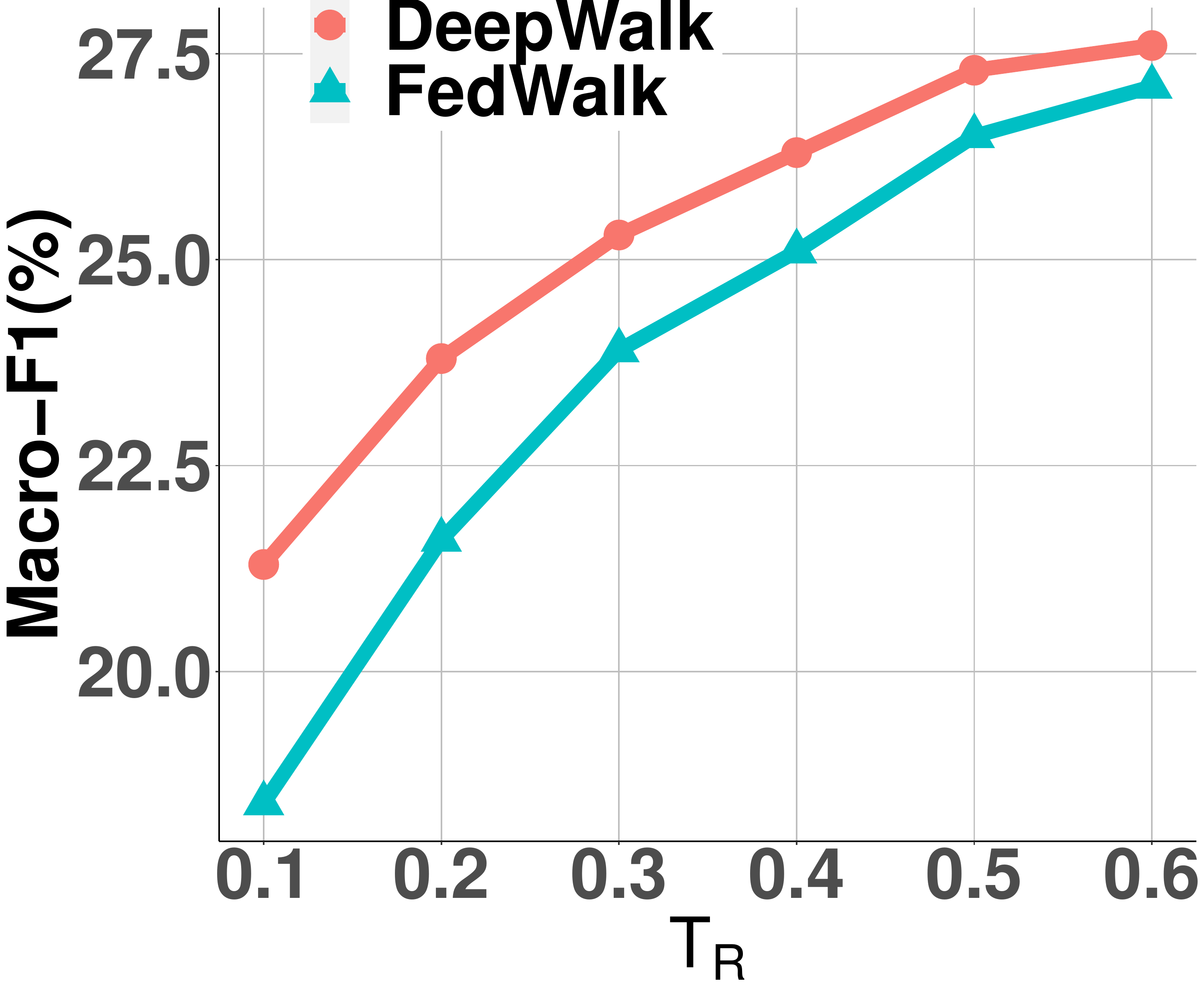}
  
    }
    \caption{Blogcatalog dataset}
    \label{fig:bc}
\end{figure}

\noindent
\textbf{Blogcatalog.}
For this dataset, we evaluate the performance of FedWalk and DeepWalk by setting the training dataset ratio $T_R$ from 0.1 to 0.6. The results are illustrated in Fig. \ref{fig:bc}. FedWalk performs close to DeepWalk does both in Micro-F1 score and in Macro-F1 score. As $T_R$ increases, the gap between these two methods closes. When $T_R=0.6$, FedWalk only loses 1.8\% Micro-F1 score and 1.0\%  Macro-F1 score compared to the scores of DeepWalk. 

\begin{figure}[h]
    \centering
    \subfigure[Micro-F1 score]
    {
        \includegraphics[width=0.22\textwidth]{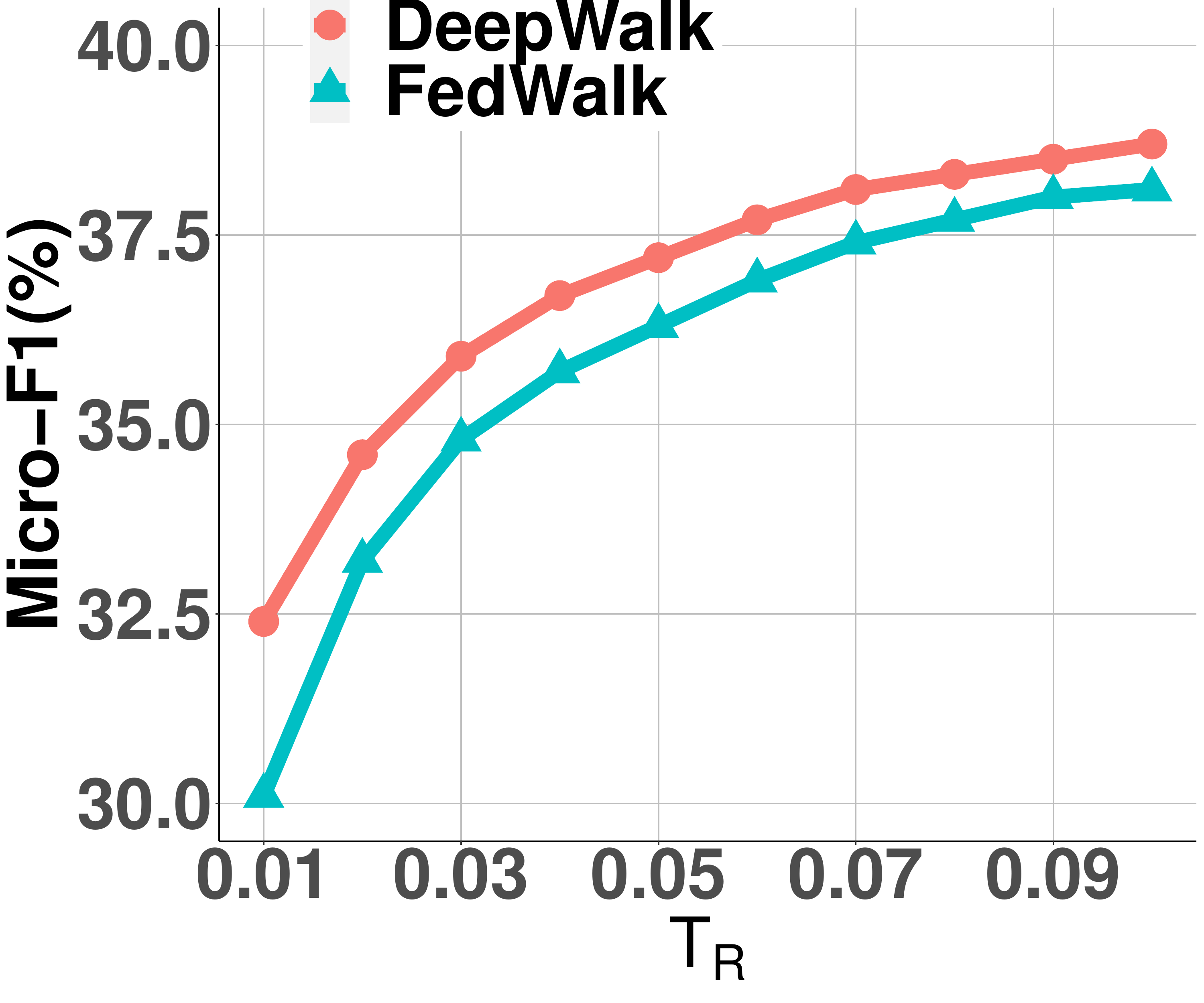}
      
    }
    \subfigure[Macro-F1 score]
    {
        \includegraphics[width=0.22\textwidth]{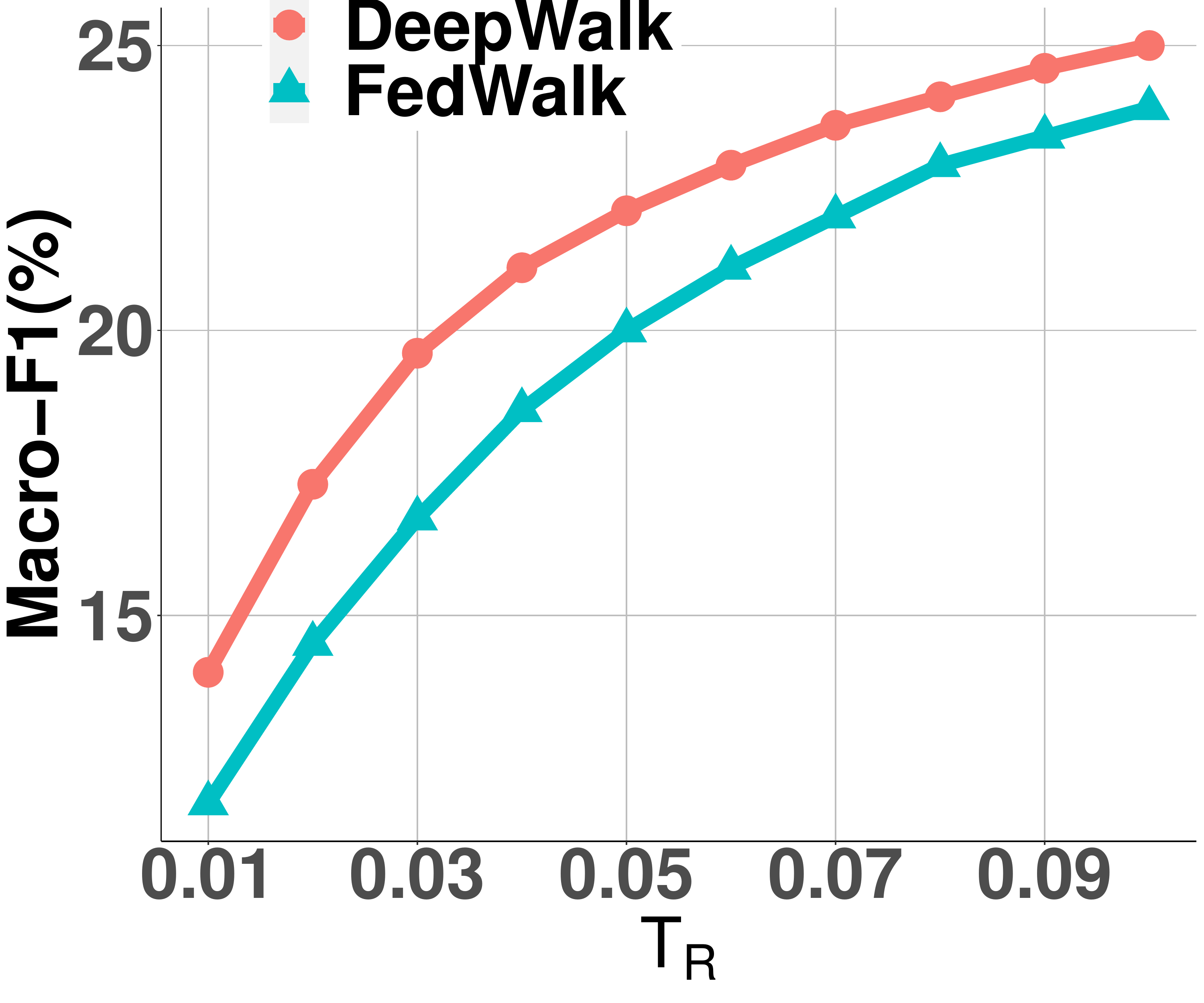}
  
    }
    \caption{Flickr dataset}
    \label{fig:fr}
\end{figure}

\noindent
\textbf{Flickr.}
For Flickr dataset, we set the training dataset ratio $T_R$ from 0.01 to 0.1 due to its large scale. The results are shown in Fig. \ref{fig:fr}. The results are consistent with those obtained from Blogcatalog dataset. Generally, DeepWalk has comparable performance to the one of FedWalk. When $T_R$ gets larger, the difference between the results of our method and the baseline method becomes smaller. When $T_R=0.1$, DeepWalk only loses 1.5\% Micro-F1 score and 4.4\% Macro-F1 score. 

The classification results of these two methods show that our method FedWalk can achieve performance very close to the centralized method DeepWalk. This implies that our method has limited utility loss in the federated regime.  
\subsection{Sensitivity Analysis}
\begin{figure}[h]
    \centering
    \subfigure[Blogcatalog dataset]
    {
        \includegraphics[width=0.22\textwidth]{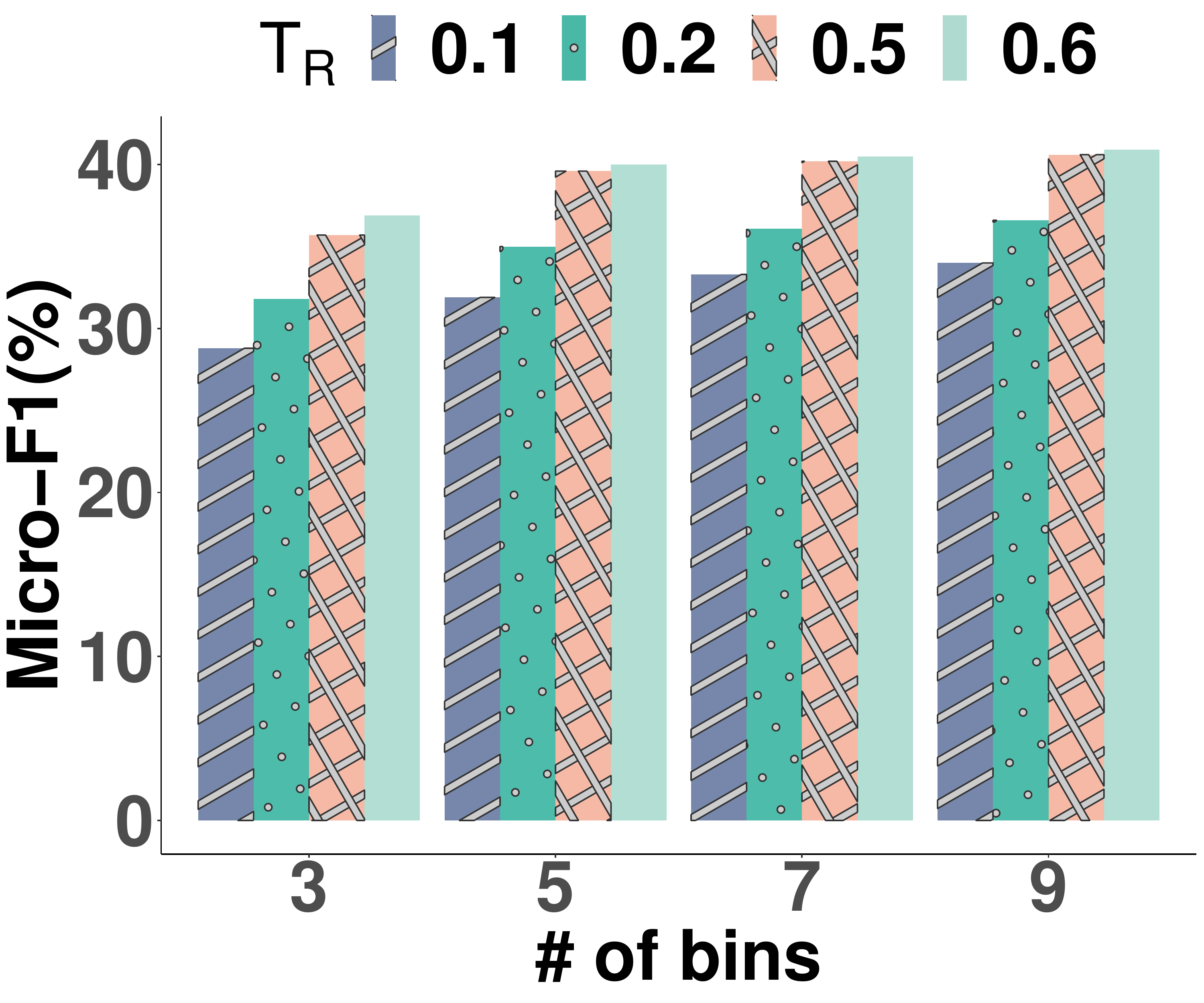}
      
    }
    \subfigure[Flickr dataset]
    {
        \includegraphics[width=0.22\textwidth]{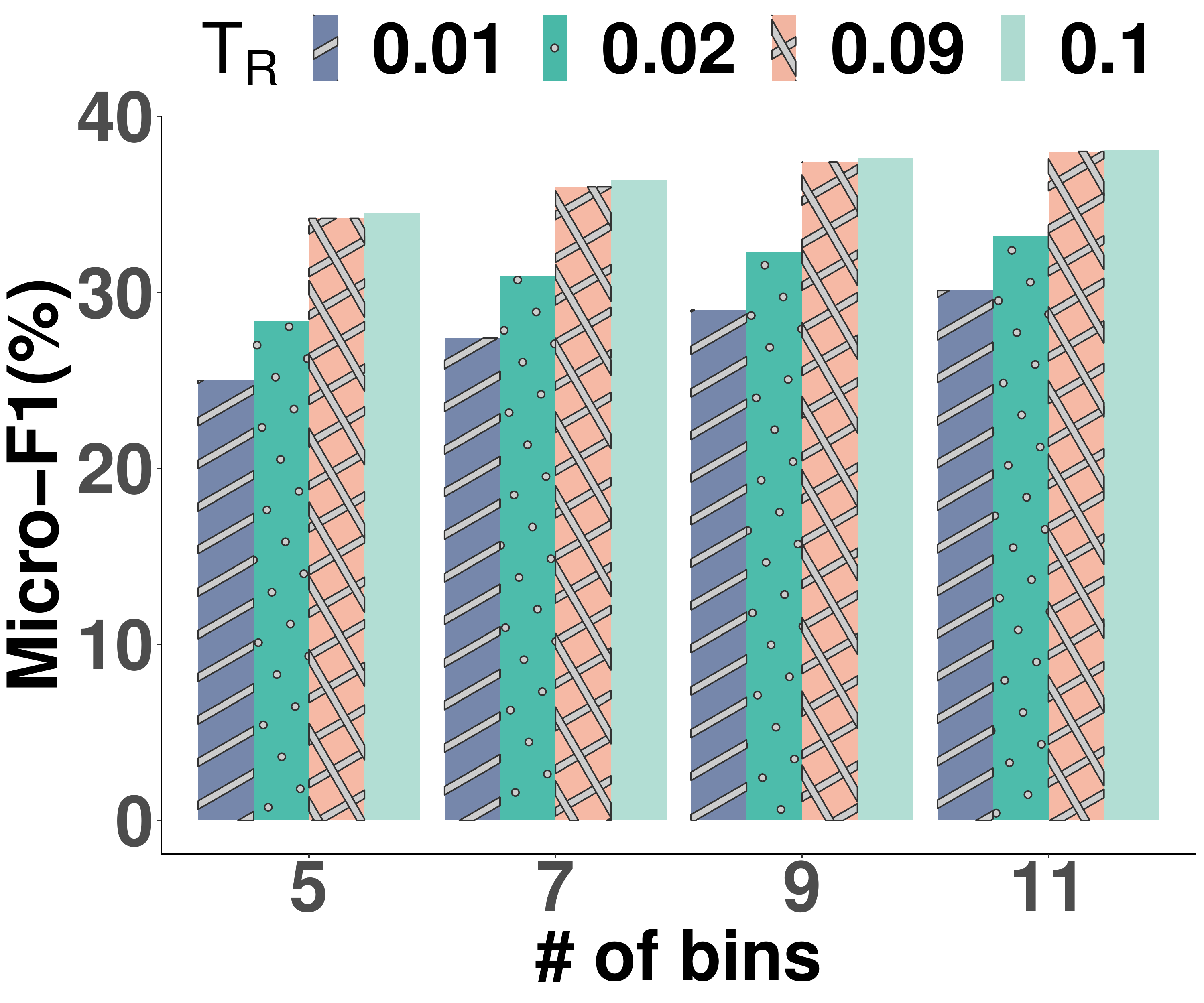}
  
    }
    \caption{Sensitivity analysis of parameter $k$}
    \label{fig:k}
\end{figure}

\textbf{Number of bins $k$.}
Figure \ref{fig:k} shows the effect of the number of bins $k$. Since Flickr Dataset has a lot more vertices than Blogcatalog Dataset, we test the performance upon different $k$ values.  For Blogcatalog dataset, increasing the number of bins from 3 to 9 results in 18.06\%, 15.09\%, 13.73\% and 10.84\% raise of Micro-F1 score respectively when $T_R=0.1,0.2,0.5$ and 0.6. For Flickr dataset, increasing the number of bins from 5 to 11 results in 20.4\%, 16.90\%, 11.11\% and 10.43\% raise of Micro-F1 score respectively when $T_R=0.01,0.02,0.09$ and 0.01. Both two bar plots show that larger $k$ leads to better performance of FedWalk model given different training ratios $T_R$. A larger number of bins generates a more representative degree vector, thus leading to a better HCT. A good HCT provides an outcome with the high utility of good random walk sequence encoder and more accurate prediction of the two-hop neighbor predictor. This produces more representative vertex embedding. 

\begin{figure}[h]
    \centering
    \subfigure[Blogcatalog dataset]
    {
        \includegraphics[width=0.22\textwidth]{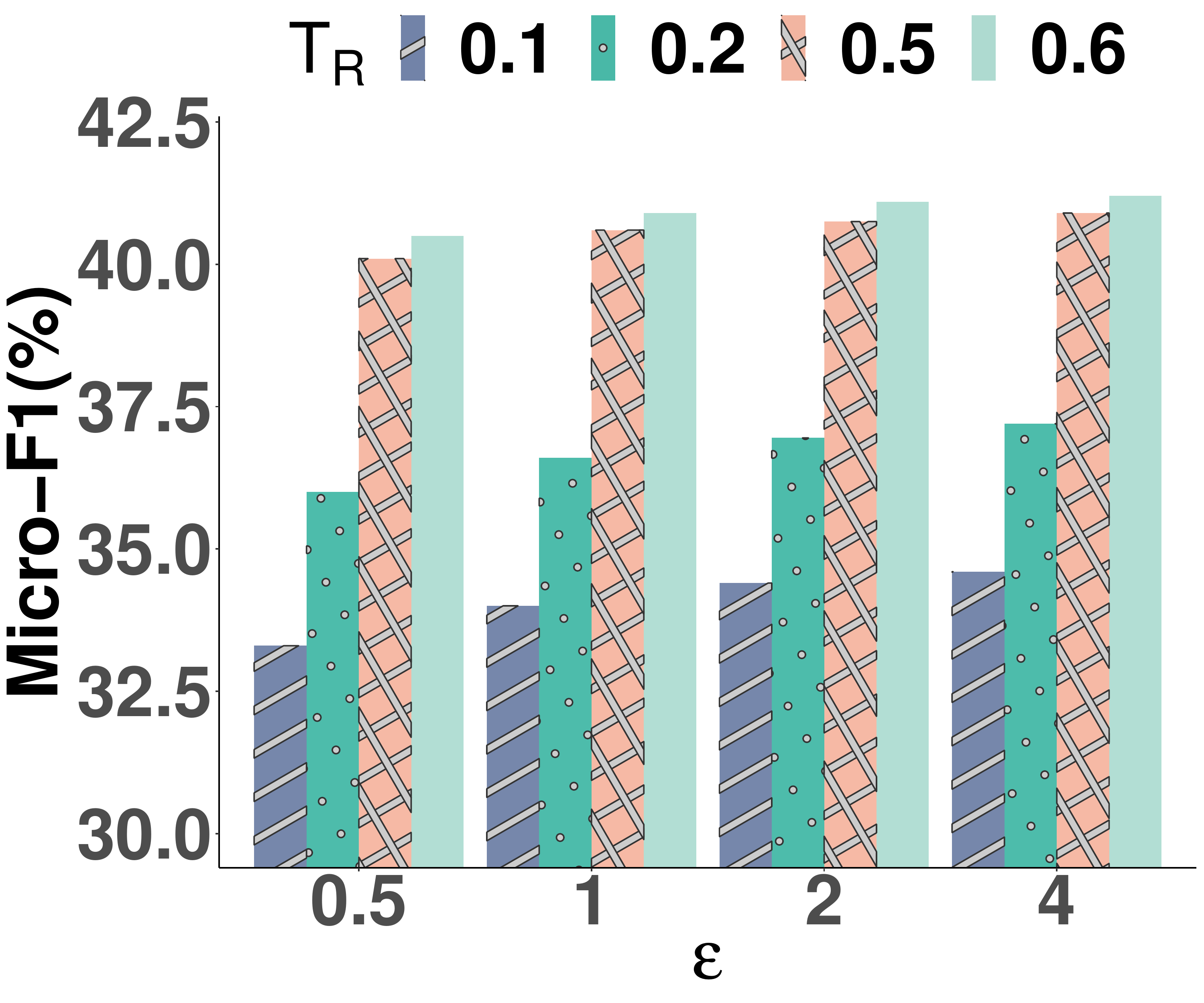}
      
    }
    \subfigure[Flickr dataset]
    {
        \includegraphics[width=0.22\textwidth]{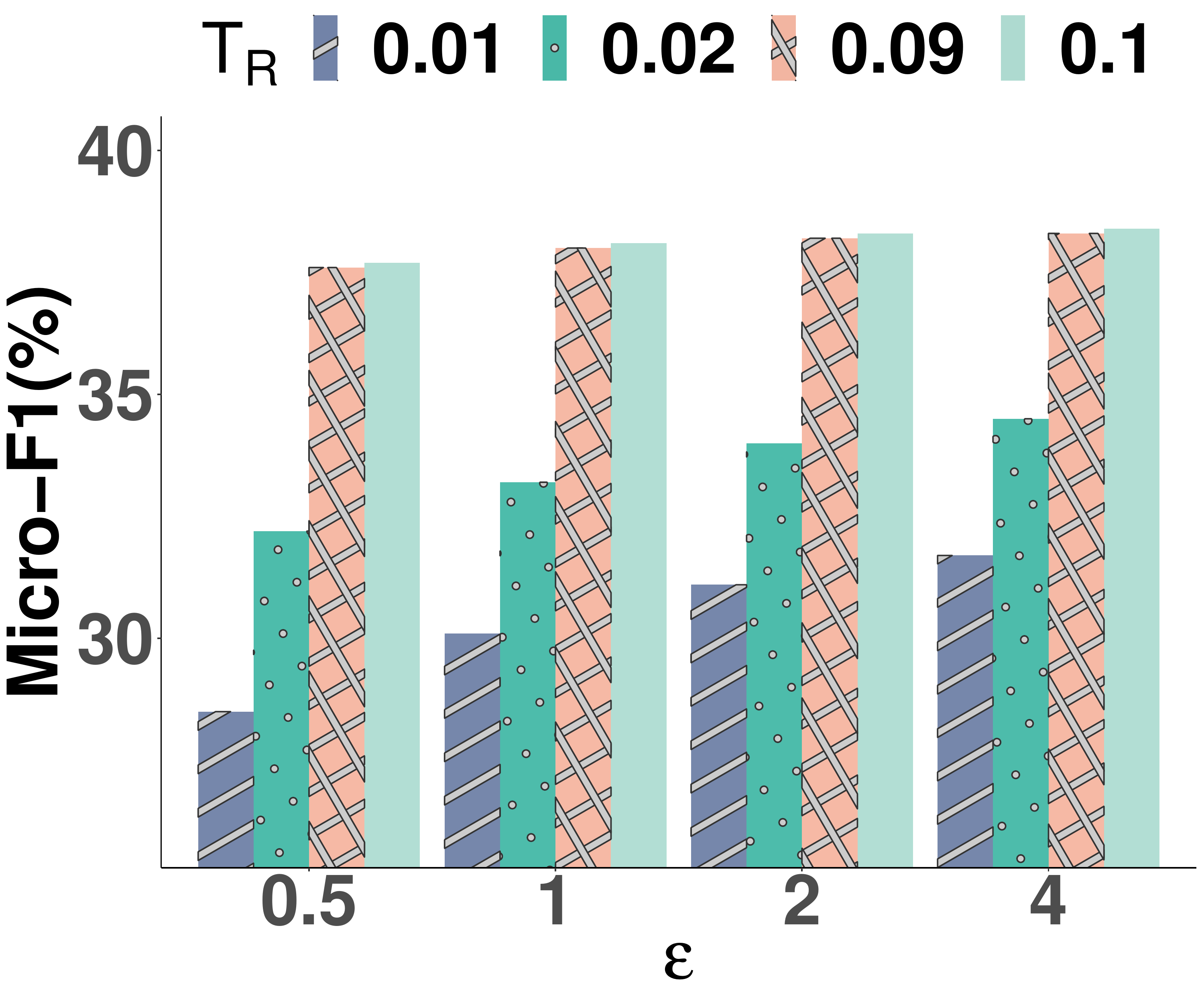}
  
    }
    \caption{Sensitivity analysis of parameter $\epsilon$}
    \label{fig:epsilon}
\end{figure}

\noindent
\textbf{Laplace noise parameter $\epsilon$.}
Figure \ref{fig:epsilon} shows the effect of Laplace noise parameter $\epsilon$ upon the performance of FedWalk model. For Blogcatalog dataset, increasing $\epsilon$ from 0.5 to 4 results in 3.90\%, 3.33\%, 2.00\% and 1.73\% raise of Micro-F1 score respectively when $T_R=0.1,0.2,0.5$ and 0.6. For Flickr dataset, the same increase of $\epsilon$ results in 11.23\%, 7.14\%, 1.86\% and 1.85\% raise of Micro-F1 score respectively when $T_R=0.01,0.02,0.09$ and 0.01. Both two plots illustrate that the model with larger $\epsilon$ performs better for different training ratios $T_R$. $\epsilon$ represents the extent to which privacy is preserved. Small $\epsilon$ protects more privacy while losing more utility. In this way, the overall performance degrades as $\epsilon$ decreases. 
\par

\begin{figure}[h]
    \centering
    \subfigure[Communication cost(Blogcatalog)]
    {
        \includegraphics[width=0.22\textwidth]{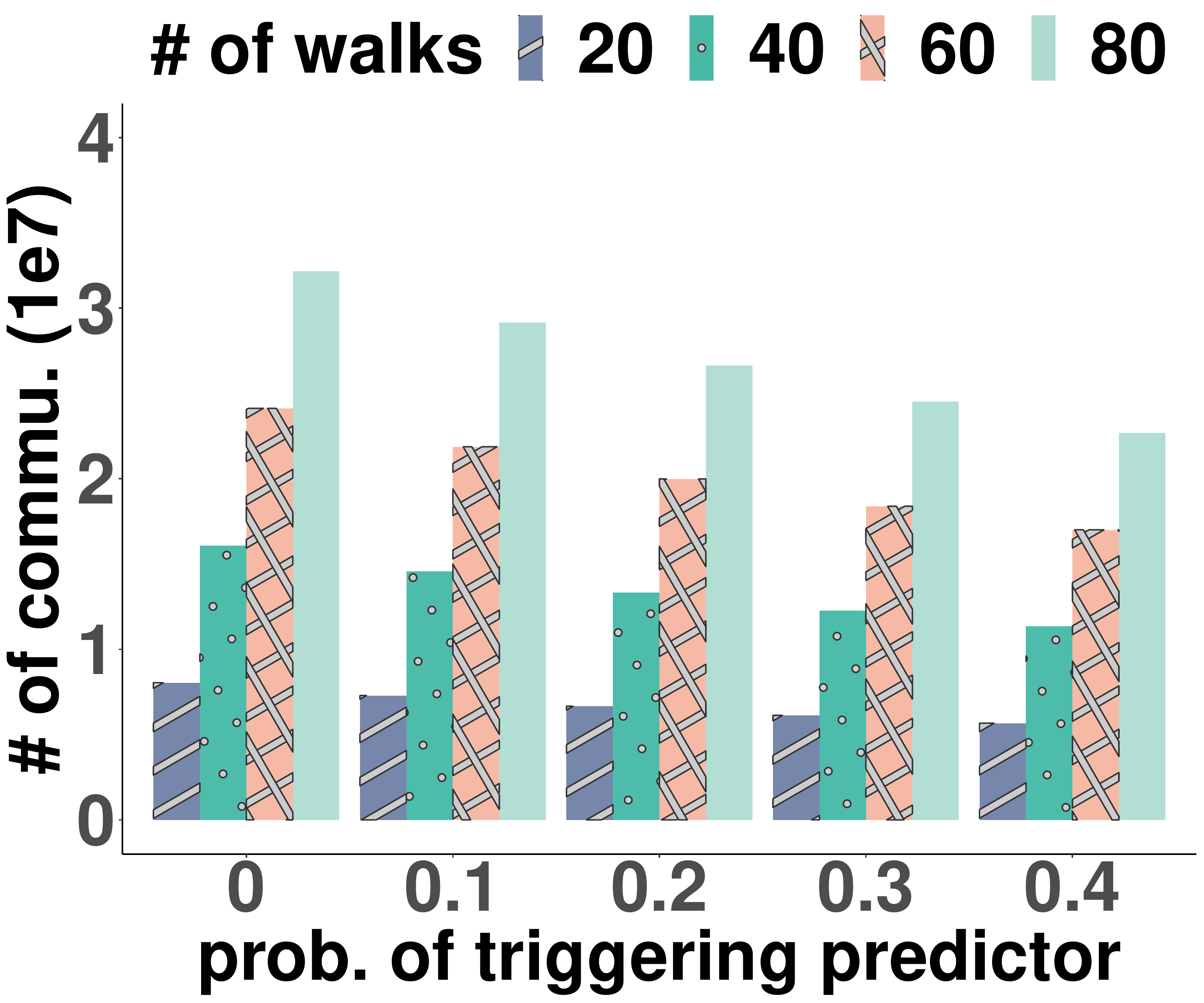}
        \label{fig:pcom}
      
    }
    \subfigure[Micro-F1 score(Blogcatalog)]
    {
        \includegraphics[width=0.22\textwidth]{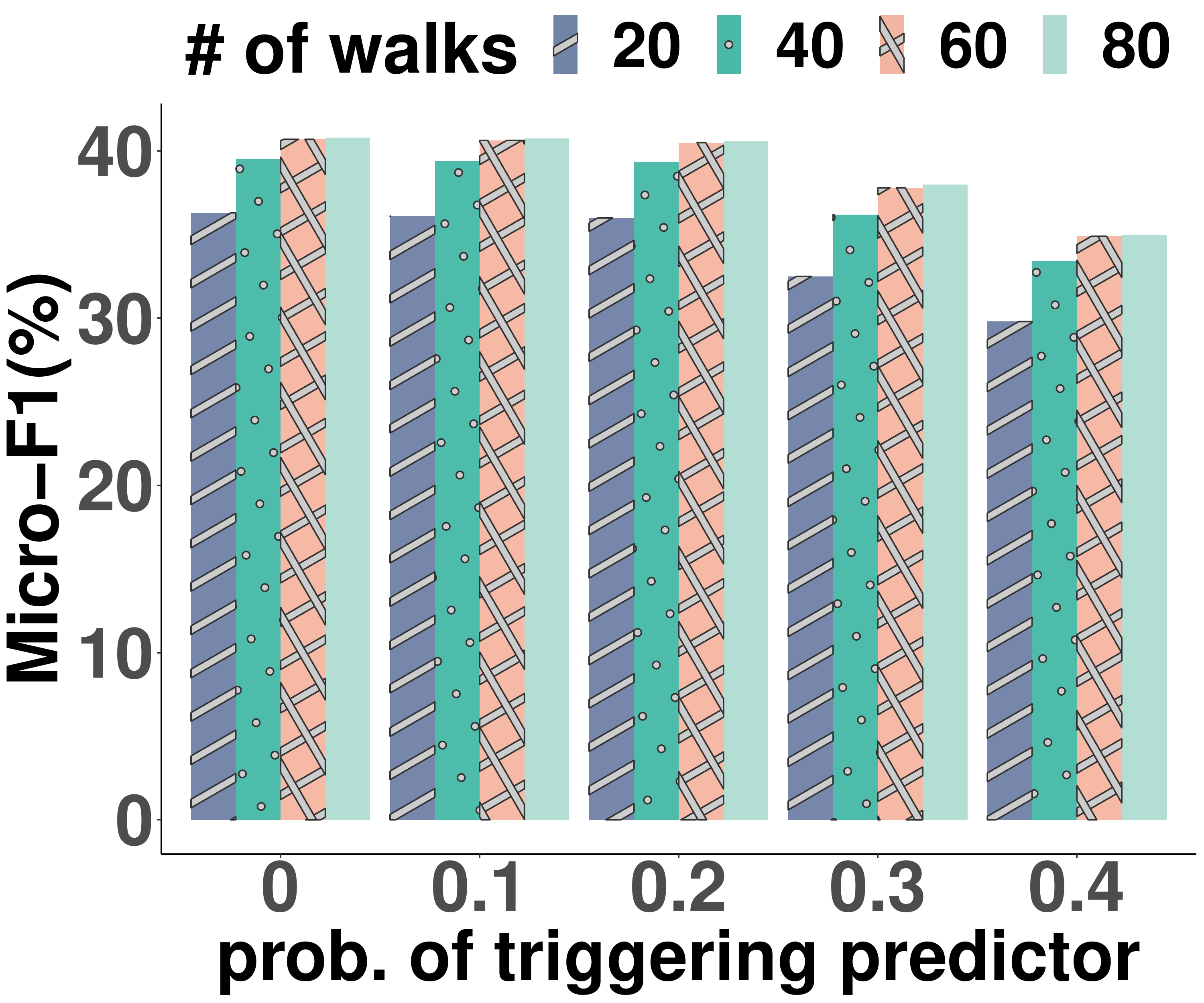}
        \label{fig:pacc}
    }
    \subfigure[Communication cost(Flickr)]
    {
        \includegraphics[width=0.22\textwidth]{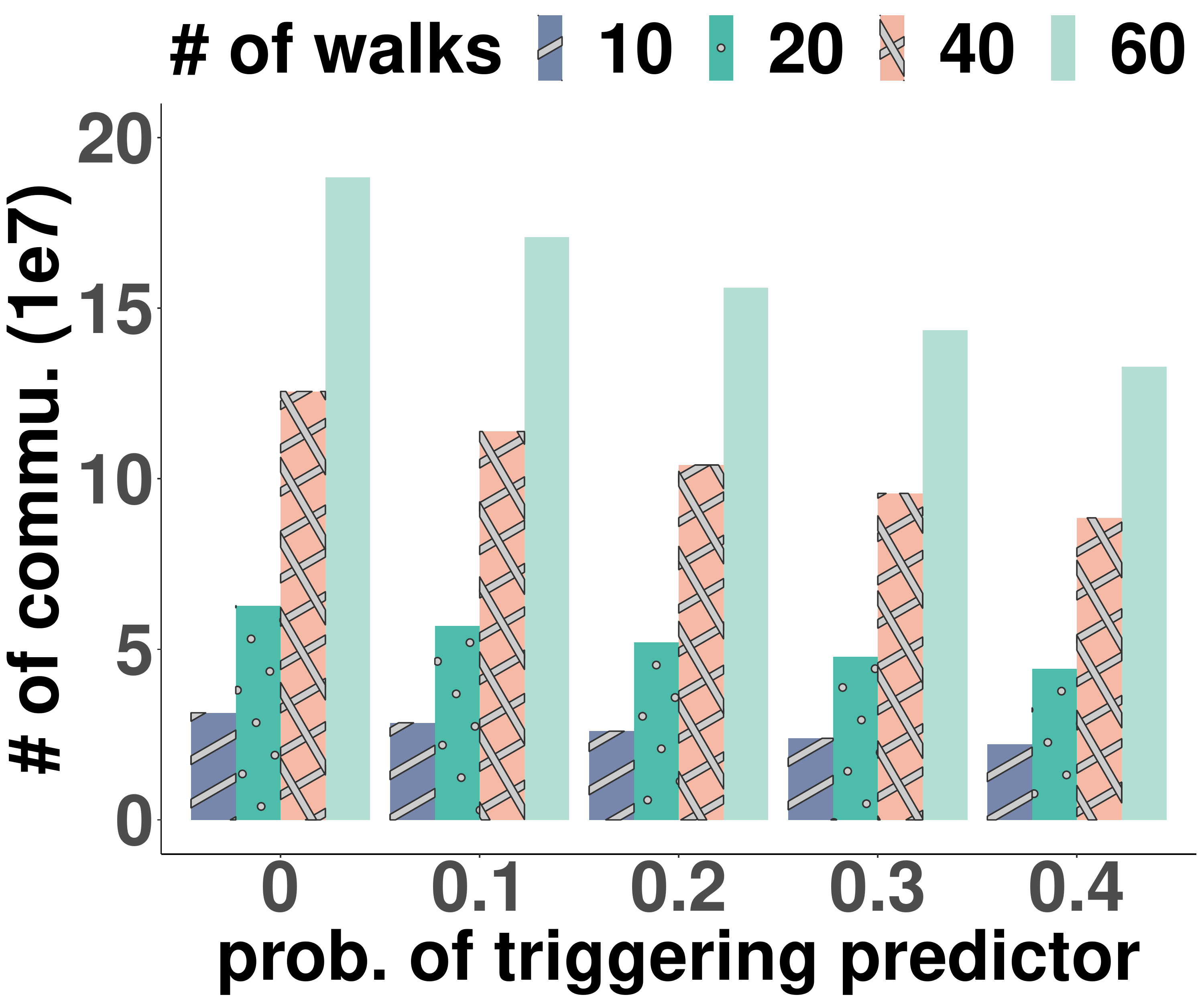}
        \label{fig:pcom2}
      
    }
    \subfigure[Micro-F1 score(Flickr)]
    {
        \includegraphics[width=0.22\textwidth]{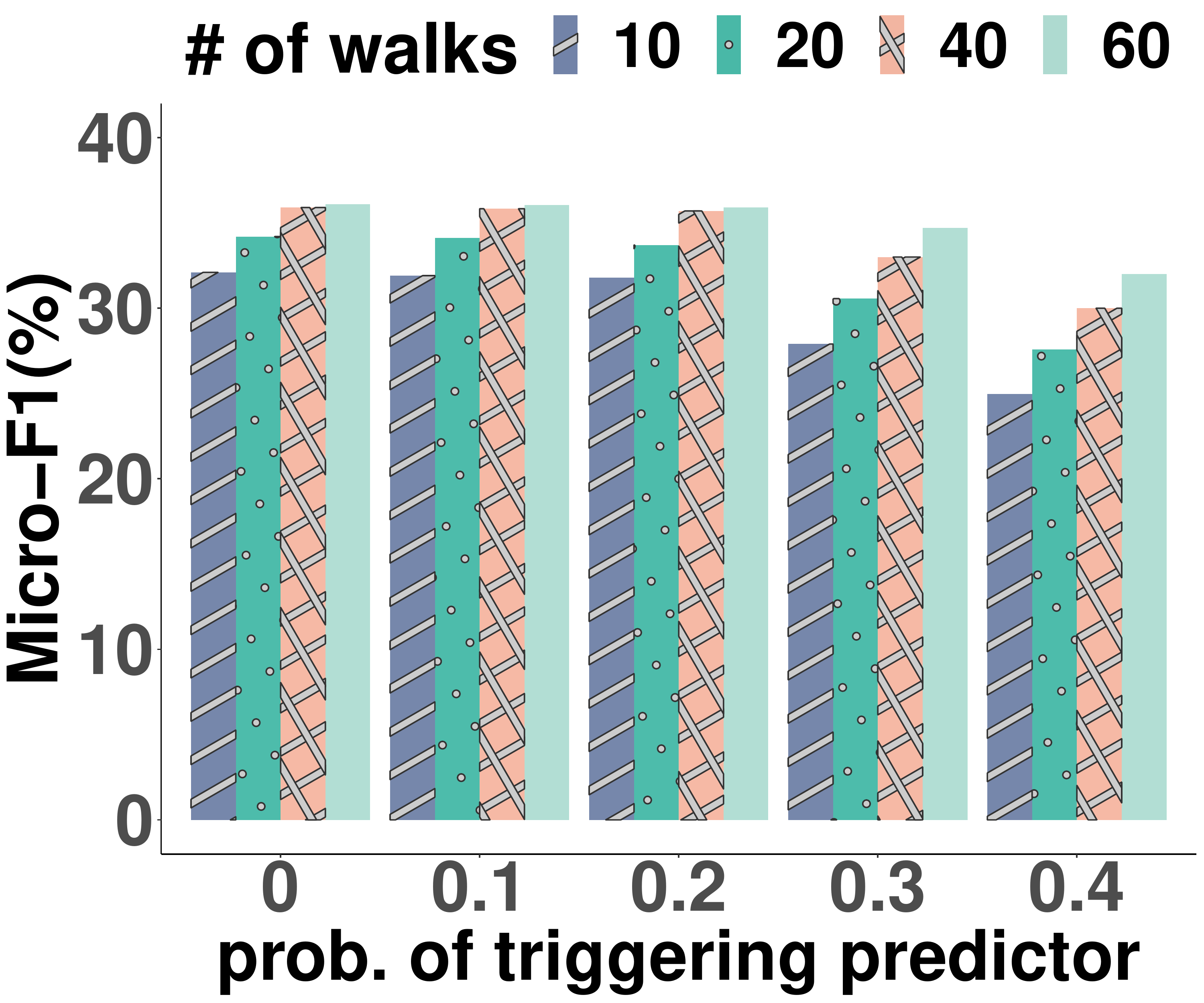}
        \label{fig:pacc2}
    }
    \caption{Sensitivity analysis of parameter $p$}
    \label{fig:p}
\end{figure}
\noindent
\textbf{Probability of triggering predictor $p$.}
We test the effect of the probability of triggering two-hop neighbor predictor $p$ given different numbers of walks per vertex. Figure \ref{fig:pcom} and Figure \ref{fig:pcom2} show the influence upon the number of inter-device communication. Increasing $p$ from 0 to 0.4 results in approximately 23.80\% decrease of the number of inter-device communication. For every 40-vertex-long random walk, it takes 39, 35.35, 32.30, 29.72 and 27.49 times of inter-device communication respectively when $p=0,0.1,0.2,0.3$ and 0.4.  The figures show that larger $p$ saves more inter-device communication as the predictor saves the communication cost by skipping the 1 hop neighbor and directly communicating to the predicted 2 hop neighbor. 

Figure \ref{fig:p} displays the trade-off relationship between communication cost and accuracy of FedWalk model. 
Figure \ref{fig:pacc} and Figure \ref{fig:pacc2} demonstrate that smaller $p$ value leads to more representative vertex embedding. Increasing $p$ from 0 to 0.4 results in 10.47\%, 8.35\%, 7.13\% and 6.86\% decrease of Micro-F1 score respectively when the numbers of walks per vertex are 20, 40, 60, and 80. For Flickr dataset, the same increase of $p$ results in 13.08\%, 10.64\%, 8.08\% and 3.88\% decrease of Micro-F1 score respectively when the numbers of walks per vertex are 10, 20, 40 and 60. Especially when $p>0.2$, the decrease of micro-F1 score is obvious.

\section{Conclusion}
\label{sec:con}
In this paper, we propose the first unsupervised node embedding algorithm in the node-level federated graphs. Our framework, FedWalk, provides centralized competitive representation capability with reduced communication cost and enhanced data privacy. In FedWalk, we first propose a differentially private HCT constructor to capture structural information. We then propose a random walk generator which includes a sequence encoder to preserve privacy and a two-hop neighbor predictor to save communication cost. 
FedWalk is theoretically proved to have bounded dissimilarity value loss and quantifiable communication reduction while preserving $\epsilon-$ differential privacy. Extensive experiments on real-world datasets validate the effectiveness and efficiency of our design.

\begin{acks}
This research is supported by SJTU Explore-X grant.
\end{acks}

\bibliographystyle{ACM-Reference-Format}
\balance
\bibliography{main.bbl}

\newpage
\appendix 
\section{Proofs}
In this section, we provide proofs for the theorems mentioned in this paper. 

\subsection{Proof of Theorem \ref{th:ub}}
\begin{proof}
Let the noised ordered degree matrix of vertex $u$ be $\bm{M_u}$ and the original ordered degree matrix of vertex $u$ be $\bm{M_u}'$ such that $M_{u,ij}=M_{u,ij}'+R_{u,ij}$. Recall that the dissimilarity value is calculated using Eq.\ref{dp} and Eq.\ref{dissim}. In this way, 
\begin{equation}
\begin{aligned}
     &\mathbb{E}[\text{dissim}(u,v)-\text{dissim}'(u,v)]\\
     =&\mathbb{E}\left[ \sum ||\bm{c_{u_i}}-\bm{c_{v_j}}||_2-\sum ||\bm{c_{u_{i'}}'}-\bm{c_{v_{j'}}'}||_2\right]
\end{aligned}
\end{equation}
where $(i,j)$ and $(i',j')$ are some pairs selected from the $[1,|N(u)|]\times [1,|N(v)|]$ integer space according to Eq.\ref{dp}. Thus, 
\begin{equation}
\begin{aligned}
     &\mathbb{E}[\text{dissim}(u,v)-\text{dissim}'(u,v)]\\
     =&\mathbb{E}\left[\sum ||\bm{c_{u_i}}-\bm{c_{v_j}}||_1-\sum ||\bm{c_{u_{i'}}'}-\bm{c_{v_{j'}}'}||_1\right]\\
     \leq &\mathbb{E}\left[\sum_{i=1}^{|N(u)|}\sum_{j=1}^{|N(v)|} ||\bm{c_{u_i}}-\bm{c_{v_j}}||_1- \sum_{i'=1}^{|N(u)|}\sum_{j'=1}^{|N(v)|} ||\bm{c_{u_{i'}}'}-\bm{c_{v_{j'}}'}||_1\right]\\
     =&\mathbb{E}\left[\sum_{i=1}^{|N(u)|}\sum_{j=1}^{|N(v)|} ||\bm{c_{u_i}}-\bm{c_{v_j}}||_1-||\bm{c_{u_{i}}'}-\bm{c_{v_{j}}'}||_1\right]\\
     = &\mathbb{E}\left[\sum_{i=1}^{|N(u)|}\sum_{j=1}^{|N(v)|} \sum_{l=1}^k |M_{u,il}-M_{v,jl}|-|M_{u,il}'-M_{v,jl}'|\right]\\
     \leq &\mathbb{E}\left[\sum_{i=1}^{|N(u)|}\sum_{j=1}^{|N(v)|} \sum_{l=1}^k |R_{u,il}-R_{v,jl}|\right]\\
\end{aligned}
\end{equation}
Since the noised item $R$ follows an i.i.d. Laplace distribution, we have 
$f_{R_{u,il}-R_{v,jl}}(x)=\frac{\epsilon}{4}\left( e^{-\epsilon|x|}+|x|\epsilon e^{-\epsilon|x|}\right)$
which implies that $\mathbb{E}_{|x|}\left[ f_{R_{u,il}-R_{v,jl}}(x)\right ]=\frac{3}{2\epsilon}$. Hence, we bound the expectation
\begin{equation}
\begin{aligned}
     &\mathbb{E}[\text{dissim}(u,v)-\text{dissim}'(u,v)]\\
     \leq &k|N(u)||N(v)|\frac{3}{2\epsilon}\\
     \leq &\frac{3k(\max_{v\in V}|N(v)|)^2}{2\epsilon}
\end{aligned}
\end{equation}
\end{proof}

\subsection{Proof of Theorem \ref{probcomm}}
\begin{proof}
Let the expected times of inter-device communication be $\mathbb{E}_l$ for one $l-$vertex-long random walk. 

When $l=1$, $\mathbb{E}_1=0$. When $l=2,$ $\mathbb{E}_2=1$. 
When $l\ge 3$,  the following relationship holds
\begin{equation}
\mathbb{E}_{l}=p(\mathbb{E}_{l-2}+1)+(1-p)(\mathbb{E}_{l-1}+1)
\end{equation}
which can be rewritten in the form of 
\begin{equation}
\mathbb{E}_{l}-\mathbb{E}_{l-1}=1-p(\mathbb{E}_{l-1}-\mathbb{E}_{l-2}).
\end{equation}
This gives us a recurrence relation of $\mathbb{E}_l-\mathbb{E}_{l-1}$, which implies that 
\begin{equation}
    \mathbb{E}_l = \frac{l-2}{1+p}+(2-2p-\frac{1}{1+p})\frac{1-(-p)^{l-2}}{1+p}.
\end{equation}
When $p=0$, it is expected to have $l-1$ times of communication per random walk sequence. Thus, it is expected to save 
\begin{equation}
    l-1-\left[ \frac{l-2}{1+p}+(2-2p-\frac{1}{1+p})\frac{1-(-p)^{l-2}}{1+p}\right]
\end{equation}
times of communication per random walk sequence. 
In total, it is expected to save 
\begin{equation}
    |V|\gamma \mathcal{A}
\end{equation}
where 
\begin{equation}
    \mathcal{A}=l-1-\left[ \frac{l-2}{1+p}+(2-2p-\frac{1}{1+p})\frac{1-(-p)^{l-2}}{1+p}\right]
\end{equation}
times of inter-device communication.
\end{proof}
\section{Pseduo Code}
\subsection{Pseduo code of HCT constructor}
\begin{algorithm}[h]
\caption{HCT constructor}
\label{hct-constr}
\begin{algorithmic}[1]
\REQUIRE noise parameter $\epsilon$, number of bins $k$, vertices $V$;
\ENSURE a hierarchical clustering tree;\\
\textbf{Server operation}\\
\STATE Group $V$ into $k$ bins randomly;\\
\FOR{each device}
\STATE Server sends the group plan to the local device;\\
\STATE Execute \textbf{Device operation 1};\\
\STATE Server receives the noised vector $\bm{c_v}$;
\ENDFOR
\STATE Collect all the vectors to a dictionary; \\
\FOR{each device}
\STATE Server sends the dictionary to the local device;\\
\STATE Execute \textbf{Device operation 2};\\
\STATE Server receives the degree matrix;
\ENDFOR
\STATE Compute a dissimilarity matrix using DTW; \\
\STATE Construct an HCT based on the dissimilarity matrix using a hierarchical clustering algorithm; \\
\quad \\
\textbf{Device operation 1}\\
\STATE Count the number of its neighbors belonging to each bins and get a $k$-dim vector;\\
\STATE Add noise $Lap(0,\frac{1}{\epsilon})$ to each element of $\bm{c_v}$;\\
\STATE Send the noised vector $\bm{c_v}$ to the server; \\
\quad \\
\textbf{Device operation 2}\\
\STATE Form its ordered degree matrix $\bm{M_v}$ according to Eq. \ref{degree_matrix};\\
\STATE Send the degree matrix to the server; \\
\end{algorithmic}

\end{algorithm}

\end{document}